\pdfoutput=1
\documentclass[PAPER, cernpreprint, atlasdraft=false, texlive=2023, UKenglish, texmf, orcidlogo]{atlasdoc}
 
\usepackage{atlaspackage}
\usepackage{atlasbiblatex}
 
\usepackage{atlasphysics}

\graphicspath{{logos/}{figures/}}
 
\usepackage{ANA-EXOT-2019-40-PAPER-defs}

\AtlasTitle{Search for periodic signals in the dielectron and diphoton invariant mass spectra using 139 fb$^{-1}$ of $pp$ collisions at $\sqrt{s}=13$ TeV with the ATLAS detector}

\AtlasAbstract{
A search for physics beyond the Standard Model inducing periodic signals in the dielectron and diphoton invariant mass spectra is presented using 139 fb$^{-1}$ of $\sqrt{s}=13$ TeV $pp$ collision data collected by the ATLAS experiment at the LHC.
Novel search techniques based on continuous wavelet transforms are used to infer the frequency of periodic signals from the invariant mass spectra and neural network classifiers are used to enhance the sensitivity to periodic resonances.
In the absence of a signal, exclusion limits are placed at the 95\% confidence level in the two-dimensional parameter space of the clockwork gravity model.
Model-independent searches for deviations from the background-only hypothesis are also performed.
}
 
\author{The ATLAS Collaboration}
 
\AtlasRefCode{EXOT-2019-40}
 
\PreprintIdNumber{CERN-EP-2023-073}

\AtlasJournalRef{JHEP 10 (2023) 079}
\AtlasDOI{10.1007/JHEP10(2023)079}

\hypersetup{pdftitle={ATLAS document},pdfauthor={The ATLAS Collaboration}}
 
\begin{document}
 
\maketitle

\section{Introduction}

Typical experimental searches for physics beyond the Standard Model (SM) at the Large Hadron Collider (LHC) search for new resonant peaks or non-resonant deviations in kinematic distributions, such as the invariant mass of two-particle systems.
However, more complicated signatures are possible and have received little attention in experimental searches so far.
One such possibility, which appears in several theoretical extensions to the SM, is a signal that gives rise to periodic structures.
This type of signal is predicted by models with small mass splittings among many resonances.
 
Extensions to the SM with these features include the continuum clockwork gravity model~\cite{Giudice:2016yja,Giudice:2017fmj}, which has an identical five-dimensional spacetime metric to the linear dilaton scenario~\cite{Antoniadis:2001sw,Antoniadis:2011qw}.
This spacetime metric also approximates the dual of Little String Theory~\cite{Berkooz:1997cq,Seiberg:1997zk}.
As the same theory can be interpreted as either a linear dilaton model or as the continuum version of a clockwork model, this theory is referred to as the Clockwork/Linear Dilaton (CW/LD) model.
These models are well motivated, as the continuum spacetime version of the clockwork gravity model can address the Higgs boson mass naturalness problem~\cite{Giudice:2016yja}.
The CW/LD model is also related to previous proposed solutions to the hierarchy problem, including theories with large extra dimensions~\cite{ArkaniHamed:1998rs,Antoniadis:1998ig} and Randall--Sundrum (RS) models where the extra dimension is warped~\cite{Randall:1999ee}.
 
The CW/LD model predicts a narrowly-spaced spectrum of resonances in mass.
The mass spectrum and couplings of this tower of Kaluza--Klein (KK) gravitons are described by two parameters, $k$ and $M_{5}$.
The parameter $k$ is a mass parameter that determines the onset of the KK graviton spectrum and $M_{5}$ is the five-dimensional (5D) reduced Planck mass, the fundamental scale of the theory.
As the cross-section for the CW/LD model approximately scales inversely with $M_{5}$, $\sigma \propto 1/M_{5}^{3}$, a limit on the signal strength in this search can be directly translated to a limit on $M_{5}$ for a given value of $k$.
 
The mass splittings predicted by the CW/LD model are generally on the order of a few percent near the onset of the graviton spectrum and eventually decrease, falling below 1\% for high graviton mode numbers~\cite{Giudice:2017fmj}.
Therefore, when looking at the invariant mass of the decay products of these resonances in experiments, the narrowly-spaced spectrum of resonances predicted by the model may appear as a long-range semi-periodic structure.
The narrow spacing of the mass spectrum requires an excellent detector resolution in order to resolve these resonances which can be satisfied by relying on two electrons or two photons in the final state.
This paper presents a search for periodic contributions to the diphoton and dielectron invariant mass spectra using 139~\ifb\ of $\sqrt{s} = 13$~\TeV \ proton--proton ($pp$) collision data recorded by the ATLAS detector from 2015 to 2018 at the LHC.
Previous searches for physics beyond the SM in high-mass diphoton final states and dielectron final states during Run 2 of the LHC using $pp$ collisions at $\sqrt{s} = 13$~\TeV were conducted by the ATLAS and CMS experiments~\cite{EXOT-2018-08,EXOT-2019-16,HIGG-2018-27,CMS-EXO-17-017,CMS-EXO-19-019}.
The CMS experiment performed a search for the same CW/LD signal using only diphoton events in a search for a non-resonant excess at high invariant masses.
The CMS result excludes values of $M_{5}$ in the range of 10~\TeV to 1~\TeV for values of $k$ in the range of 0.1~\GeV to 5~\TeV~\cite{CMS-EXO-17-017}.


\section{Analysis strategy}

This analysis is based on two previous ATLAS resonance searches performed in the dilepton~\cite{EXOT-2018-08} and diphoton~\cite{HIGG-2018-27} channels in the mass range of $225$--$6000$~\GeV and $150$--$5000$~\GeV, respectively.
The same data samples and background estimation methods were used, in both cases, relying on data-driven techniques.
The two previous analyses also estimated the background from simulation and found excellent compatibility with the estimations from data.
Therefore, this analysis takes advantage of the precise data-driven background estimates while also benefitting from the availability of a simulation-based background estimates.
The background estimates from simulation include a complete set of systematic uncertainties in both of the analysis channels, which is important for this analysis because the ``spurious-signal'' uncertainty derived in Refs.~\cite{EXOT-2018-08,HIGG-2018-27} is applicable only to searches which consider a single narrow resonance.
The simulation-based uncertainties from previous searches are used instead and their effect on the background estimation is incorporated using the methodology developed in a previous ATLAS search for non-resonant signatures in the dilepton channel~\cite{EXOT-2019-16}.
Finally, this analysis also relies on transformational and statistical methods developed in Ref.~\cite{Beauchesne:2019tpx} in order to deal with data featuring periodic structures.
 
A brief description of the ATLAS detector is given in Section~\ref{sec:detector}.
The data and simulated event samples are discussed in Section~\ref{sec:data} and an overview of the event selection in each analysis channel is described in Section~\ref{sec:evtsel}.
The signal modelling is presented in Section~\ref{sec:sigmod} and the background estimation methods are discussed briefly in Section~\ref{sec:backgroundmodeling}.
The sources of systematic uncertainties are discussed in Section~\ref{sec:uncertainties}, while the uncertainty estimation process is described in Section~\ref{sec:toys}.
 
This analysis performs a search for generic periodic features in the dielectron and diphoton invariant mass spectra, in the same mass ranges considered in Refs.~\cite{EXOT-2018-08,HIGG-2018-27}.
The dielectron and diphoton channels are analysed separately due to potential overlaps in their event selections.
This search uses a continuous wavelet transform (CWT) to analyse these mass spectra in the frequency domain.\footnote{In this analysis, the transformed variable is the invariant mass rather than time.}
The output of the CWT is a two-dimensional image of the wavelet amplitude in the frequency versus mass space, referred to as a ``scalogram.''
The potential periodicity of a signal can be revealed in the image, for example as a local ``blob'' around a small range of frequencies or masses.
While conventional resonant and non-resonant searches may be suboptimal, particularly in the case of small CW/LD-like signals, the CWT may still provide an enhancement of a new signal's semi-periodic nature.
This enhancement provides a clear separation of a signal from the background and suggests that the search should be conducted in the CWT space rather than in the mass space alone.
The CWT method is described briefly in Section~\ref{sSec:CWT}.

In the limit of infinite statistical precision, the local features of a potential signal in the CWT images can be clearly visible and separable from the background-only case.
When accounting for realistic statistical fluctuations due to the finite-size distributions in the mass space, the signal features wash out partially in the CWT images and the separation power becomes significantly smaller.
Therefore, machine-learning techniques are applied to distinguish periodic contributions to the diphoton and dielectron CWT images made using the invariant mass spectra.
Model-independent results are provided using an autoencoder-based anomaly detection procedure to search for generic periodic deviations in the scalograms.
The CW/LD model is used as a benchmark model for a model-specific search using the CWT images.
A neural network binary classifier is trained on the CWT images made from background-only and signal-plus background mass distributions to provide a test statistic for discovery of potential periodic resonances with specific $k$ and $M_{5}$ values.
In both the model-independent and model-specific searches, the test-statistic used during the statistical inference portion of the analysis is based on the machine-learning outputs themselves.
As the CW/LD signal may introduce a non-resonant enhancement above the SM background in the mass space, the statistical inference in the two searches can be dominated by the non-resonant signal features instead of the semi-periodic contribution.
Therefore, the two searches are performed with and without specific thresholds imposed to reduce potential non-resonant contributions and focus the sensitivity on the periodic features.
The statistical analysis of the two methods and the related thresholds are discussed in Section~\ref{sec:statanalysis}.
The model-independent results and the model-specific exclusion limits set in the $k$--$M_{5}$ plane of the CW/LD model are given in Section~\ref{sec:result}.
 
The previous dilepton analysis~\cite{EXOT-2018-08} also included a dimuon analysis channel that could be sensitive to a potential CW/LD signal.
However, the mass resolution becomes much worse for high-mass dimuon events compared with the dielectron channel.
These resolution effects cause the individual modes in the KK tower to merge and the signal periodicity cannot be resolved.
For this reason, the dimuon channel is not used in this search.


\section{ATLAS detector}
\label{sec:detector}

The ATLAS experiment~\cite{PERF-2007-01,ATLAS-TDR-2010-19,PIX-2018-001} is a multipurpose particle detector with a forward-backward symmetric cylindrical geometry and nearly 4$\pi$ coverage in solid angle.\footnote{ATLAS uses a right-handed coordinate system with its origin at the nominal interaction point (IP) in the centre of the detector and the $z$-axis along the beam pipe. The $x$-axis points from the IP to the centre of the LHC ring, and the $y$-axis points upward. Cylindrical coordinates $(r,\phi)$ are used in the transverse plane, $\phi$ being the azimuthal angle around the $z$-axis. The pseudorapidity is defined in terms of the polar angle $\theta$ as $\eta=-\ln\tan(\theta/2)$.}
It consists of an inner tracking detector surrounded by a thin superconducting solenoid providing a $2$T axial magnetic field, electromagnetic and hadronic calorimeters, and a muon spectrometer.
The inner detector (ID) covers the pseudorapidity range $|\eta| < 2.5$ and consists of silicon pixel, silicon microstrip, and transition radiation tracking detectors.
Lead/liquid-argon (LAr) sampling calorimeters provide electromagnetic (EM) energy measurements with high granularity.
A steel/scintillating-tile hadron sampling calorimeter covers the central pseudorapidity range ($|\eta| < 1.7$).
The endcap and forward regions are instrumented with LAr calorimeters for both the EM and hadronic energy measurements up to $|\eta| = 4.9$.
The muon spectrometer surrounds the calorimeters, covering the region $|\eta| < 2.7$ and is based on three large superconducting air-core toroidal magnets with eight coils each.
The field integral of the toroids ranges between 2.0 and 6.0 Tm across most of the detector acceptance.
A two-level trigger system is used to select events.
The first-level trigger is implemented in custom hardware and uses a subset of the detector information to accept events at a rate below \SI{100}{\kHz}.
A software-based trigger then reduces the accepted event rate to an average of 1 kHz for offline storage~\cite{TRIG-2016-01}.
An extensive software suite~\cite{ATL-SOFT-PUB-2021-001} is used in the reconstruction and analysis of real and simulated data, in detector operations, and in the trigger and data acquisition systems of the experiment.


\section{Data and simulated event samples}
\label{sec:data}

The data  used for the search consists of the \pp collision data recorded by the ATLAS experiment at $\sqrt{s}=13$~\TeV\ during the 2015 to 2018 LHC data-taking period.
After requiring stable beam conditions and data quality selections with all ATLAS subsystems operational~\cite{DAPR-2018-01}, the data sample corresponds to an integrated luminosity of 139.0 $\pm$ 2.4~\ifb~\cite{ATLAS-CONF-2019-021}.
The LUCID-2 detector~\cite{Avoni:2018iuv} was used for the primary luminosity measurements.
 
The events used in this analysis were recorded using a set of diphoton and dielectron triggers.
Events in the diphoton channel were recorded using a diphoton trigger that required at least two energy clusters in the EM calorimeter with transverse energies (\ET) greater than 35 and 25~\GeV for the \ET-ordered leading and subleading photon candidates, respectively.
Both of the clusters were required to satisfy  photon identification criteria based on the shower shapes in the EM calorimeters.
The triggers used in 2015 and 2016 required both of the photons to satisfy the {\textit{loose}} identification requirement~\cite{EGAM-2018-01}.
In 2017 and 2018, due to the greater instantaneous luminosity, the diphoton trigger requirements were tightened and both of the photons were required to satisfy the {\textit{medium}} identification requirement.
The efficiency of the diphoton trigger relative to the event selection given in Section~\ref{sec:evtsel} is over 99\% for the 2015--2016 data and above 98\% for the 2017--2018 data~\cite{TRIG-2018-05}.
 
Events in the dielectron channel were collected using several dielectron triggers.
The trigger used in 2015 required both of the electrons to satisfy the {\textit{loose}} identification criteria and \ET\ thresholds of 12~\GeV.
In 2016, the \ET\ thresholds were increased to 17~\GeV for both electrons and the electrons were required to satisfy  the {\textit{very loose}} identification criteria~\cite{EGAM-2018-01}.
In 2017 and 2018, the identification criteria were left unchanged and the \ET\ thresholds were increased to 24~\GeV for both of the electrons~\cite{TRIG-2018-05}.
 
Although the background in this analysis is estimated by using data-driven methods, simulated Monte Carlo (MC) events are used to optimise the analysis selections, determine appropriate fit functions for the data, estimate background compositions, and evaluate signal acceptances and efficiencies.
As the KK modes in the CW/LD model are on-shell, the interference effects between the resonant signals and all background processes are neglected in both the diphoton and dielectron channels.
 
For the diphoton channel, the largest background comes from the production of two prompt photons which represents the irreducible background in this search channel.
Smaller background contributions come from events containing a photon and a jet and events with two jets, where the jets are misidentified as photons.
These smaller backgrounds are estimated by using a data-driven technique, the two-dimensional sideband method, described in Ref.~\cite{STDM-2012-05}.
 
Events with two prompt photons were simulated using the \Sherpa[2.2.4]~\cite{Gleisberg:2008ta, Bothmann:2019yzt} event generator.
Matrix elements were calculated with up to one additional parton at next-to-leading-order (\NLO) and including two or three additional partons at leading-order (\LO) in perturbative QCD (pQCD).
These matrix elements were merged with the \SHERPA parton-shower simulation using the ME+PS@NLO prescription~\cite{Hoeche:2011fd,Hoeche:2012yf,Catani:2001cc,Hoeche:2009rj}.
The \NNPDF[3.0nnlo] parton distribution function (PDF) set~\cite{Ball:2014uwa} was used and paired with a dedicated parton-shower tune in the \SHERPA generator.
 
For the dielectron channel, the main prompt backgrounds arise from Drell--Yan (DY), top-quark pair (\ttbar), single-top-quark, and diboson production.
The background contribution from non-prompt electrons from multijet and \Wjets processes is estimated by using a data-driven technique, the matrix method, as described in Ref.~\cite{EXOT-2016-05}.
 
The DY~\cite{ATL-PHYS-PUB-2016-003} sample was generated using the \POWHEGBOX[v1]~\cite{Nason:2004rx,Frixione:2007vw,Alioli:2010xd,Alioli:2008gx} event generator with the CT10 PDF set~\cite{Lai:2010vv} and interfaced with the \Pythia[8.186]~\cite{Sjostrand:2007gs} parton shower program.
Next-to-next-to-leading-order (\NNLO) corrections in pQCD and \NLO corrections in electroweak (EW) theory were calculated and applied to the simulated DY events.
The pQCD corrections were computed with VRAP v0.9~\cite{Anastasiou:2003ds} and the CT14 \NNLO PDF set~\cite{Dulat:2015mca}.
The EW corrections were computed with MCSANC~\cite{Arbuzov:2015yja} which accounts for quantum electrodynamic effects due to initial-state radiation, interference between initial and final-state radiation, and Sudakov logarithm single-loop corrections.
 
The diboson~\cite{ATL-PHYS-PUB-2016-002} processes with fully leptonic and semileptonic final states were simulated using \Sherpa[2.2.1] with the CT10 PDF set.
Matrix elements were calculated at \NLO accuracy in QCD for up to one additional parton and at LO accuracy for up to three additional parton emissions.
The matrix element calculations were matched and merged with the \SHERPA parton shower based on Catani--Seymour dipole factorisation~\cite{Gleisberg:2008fv,Schumann:2007mg} using the ME+PS@NLO prescription.
The diboson and DY backgrounds were generated in slices of dilepton mass in order to enhance the MC statistical precision in the high-mass region.
 
The $t\bar{t}$ and single-top-quark samples were generated with \POWHEGBOX[v2]~\cite{Frixione:2007nw,Alioli:2009je,Frederix:2012dh,Re:2010bp,Nason:2004rx,Frixione:2007vw,Alioli:2010xd} at \NLO using the \NNPDF[3.0nlo] PDF~\cite{Ball:2014uwa} in the matrix element and interfaced to \Pythia[8.230]~\cite{Sjostrand:2014zea} in order to model the parton shower, hadronisation, and underlying event, with parameters set according to the A14 tune~\cite{ATL-PHYS-PUB-2014-021} and using the \NNPDF[2.3lo] set of PDFs~\cite{Ball:2012cx}.
These samples were normalised to the theoretical cross-sections calculated at \NNLO in pQCD and include resummation of the next-to-next-to-leading logarithmic soft gluon terms as provided by \TOPpp[2.0]~\cite{Czakon:2011xx}.
 
Resonant single-graviton MC samples were simulated using a Randall--Sundrum model~\cite{Randall:1999ee}.
These MC samples were generated at LO in pQCD using \Pythia[8] with the \NNPDF[2.3lo] PDF set and the A14 tune.
In these samples, only the lightest KK graviton excitation was generated.
For the diphoton decay channel, the samples were generated with a KK graviton mass $m_{G^{*}}$ in the range of 150--5000~\GeV.
A fixed coupling value of $\kMpl = 0.01$, where $\overline{M}_{\mathrm{Pl}} = M_{\mathrm{Pl}}/\sqrt{8\pi}$ is the reduced Planck scale, was used to ensure a sufficiently narrow-width signal.
For the dielectron decay channel, samples were generated with masses in the range of 200--6000~\GeV and with a fixed coupling value of $\kMpl = 0.1$.
These samples are used to determine the acceptance and efficiency of selecting the CW/LD signal.
 
Additional MC samples incorporating a series of RS graviton resonances were generated at LO in pQCD using \Pythia[8.244] with the \NNPDF[2.3lo] PDF set and the A14 tune.
In these samples, only the direct graviton decays into dielectron and diphoton final-states were simulated.
Samples were generated for $k$ values of 300~\GeV, 500~\GeV, 1~\TeV, and 2~\TeV with $M_{5}$ fixed to a value of 6~\TeV.
These samples are used to validate the analytic signal templates of the CW/LD signals, which are discussed in detail in Section~\ref{sec:sigmod}.
 
The effects of multiple \pp interactions in the same bunch crossing as that of the hard scatter plus those from adjacent bunch crossings (pile-up) are included in all simulated samples.
Pile-up collisions were generated with \Pythia[8.186] using the \NNPDF[2.3lo] PDF set and the ATLAS A3 set of tuned parameters~\cite{ATL-PHYS-PUB-2016-017}.
Simulated event samples were weighted to reproduce the distribution of the average number of interactions per bunch crossing observed in the data~\cite{STDM-2015-05}.
 
The spin-2 diphoton simulated signal events, and DY, \ttbar, single-top-quark, and diboson background events were processed using a detailed simulation of the ATLAS detector~\cite{SOFT-2010-01} based on \GEANT~\cite{Agostinelli:2002hh}.
The irreducible prompt $\gamma\gamma$ background and spin-2 dielectron signals were processed using a fast simulation of the ATLAS detector~\cite{ATL-PHYS-PUB-2010-013}, where the full simulation of the calorimeter is replaced with a fast parameterisation of the calorimeter response.
All simulated events were reconstructed with the same reconstruction algorithms as those used for data.
 
Generator-level-only MC samples of the \NLO DY background are used for the modelling studies described in Section~\ref{sec:backgroundmodeling}.
These samples could not be produced with the ATLAS detector simulation due to the large number of events required~\cite{EXOT-2018-08}.


\section{Object and event selection}
\label{sec:evtsel}

Complete descriptions of the object definitions and event selections are given in Ref.~\cite{EXOT-2018-08} for the dielectron channel and Ref.~\cite{HIGG-2018-27} for the diphoton channel.
The criteria are identical to the ones applied in this analysis and a brief description is given here.
 
The event selection in the diphoton channel requires at least two reconstructed photon candidates with $\ET > 25$~\GeV and $|\eta| < 2.37$, excluding candidates in the transition region $1.37 < |\eta| < 1.52$ between the barrel and endcap EM calorimeters.
The two highest-$E_{\textrm{T}}$ photons are used to form the diphoton candidate and are used with additional information from the tracking detectors to identify the diphoton primary vertex~\cite{HIGG-2013-08}.
After the diphoton primary vertex is identified, the leading and subleading photons are required to have $E_{\textrm{T}}/m_{\gamma\gamma} > 0.35$ and 0.25, respectively.
The diphoton invariant mass, $m_{\gamma\gamma}$, is required to be greater than 150~\GeV.
To reduce the background from jets, photon candidates must satisfy the {\textit{tight}} identification criteria based on shower shapes in the EM calorimeter~\cite{EGAM-2018-01}.
 
Events in the dielectron channel are selected by requiring at least one pair of reconstructed electron candidates. Each event is required to contain at least one reconstructed $pp$ interaction vertex, where the primary vertex is defined as the vertex with the highest sum of track transverse momenta squared.
Electron candidates are reconstructed from ID tracks matched to energy clusters deposited in the EM calorimeter~\cite{ATL-PHYS-PUB-2017-022,EGAM-2018-01}.
All selected electrons are required to have $E_{\textrm{T}} > 30$~\GeV\ and $|\eta| < 2.47$.
As in the diphoton channel, electron candidates in the transition region ($1.37 < |\eta| < 1.52$) are not considered. The final selection requires that the electrons satisfy the {\textit{medium}} identification working point.
 
Electron candidate tracks are required to satisfy $|d_{0}/\sigma(d_{0})| < 5$ and $|z_{0} \, \textrm{sin} \, \theta| < 0.5$ mm, where $d_{0}$ and $z_{0}$ are the transverse and longitudinal impact parameters defined relative to the primary vertex position and $\sigma(d_{0})$ is the uncertainty on $d_{0}$.
This selection ensures that the electron candidate track is consistent with the primary vertex of the event.
 
If there are more than two electrons in the same event, the two electrons with the highest $E_{\textrm{T}}$ are selected to form the dielectron pair.
An opposite-sign electron charge requirement is not applied due to a high probability of charge misidentification for high-$E_{\textrm{T}}$ electrons.
The reconstructed invariant mass of the dielectron pair, $m_{ee}$, is required to be above 225 GeV to avoid the region dominated by decays of the $Z$ boson which cannot be described using the same background parameterisation as the high-mass region.

To suppress the backgrounds from misidentified jets, the photon and electron candidates are required to satisfy calorimeter- and track-based isolation criteria.
The photons are required to satisfy the \textit{tight}  isolation which is 90--95\% efficient for the analysis selection~\cite{EGAM-2018-01}.
Electrons are required to satisfy the \textit{gradient} isolation which is 99\% efficient for the analysis selection~\cite{EGAM-2018-01}.


\section{Signal modelling}
\label{sec:sigmod}

The CW/LD signals are modelled using analytic invariant mass templates constructed from a range of inputs including PDF information, branching ratio, and detector resolution modelling derived from the previous ATLAS diphoton and dielectron analyses.
 
For each reconstructed mass bin, the expected number of signal events is determined by computing the contribution from the entire KK graviton tower.
The estimate of the expected number of signal events for a given reconstructed mass bin $i$ is given by
\begin{equation}
N_i^\textrm{reco} = L_\textrm{int} \cdot \sum_{G}\sigma_G \cdot{\cal B}_G (G\to XX)\cdot P(M_{i}|M_{G}) \cdot (A\times\epsilon)_{G}
\label{eq:grav_signal_modeling}
\end{equation}
where the index $G$ runs over the KK graviton modes whose masses are within the search range.
In \cref{eq:grav_signal_modeling}, $L_\textrm{int}$ is the integrated luminosity, $\sigma_G$ is the production cross-section, ${\cal B}_{G}$ denotes the branching ratio, and $(A \times \epsilon)_{G}$ is the acceptance times efficiency for a given mass, $M_{G}$, of the KK graviton.
The $P(M_{i}|M_{G})$ term represents the probability to reconstruct an event in invariant mass bin $i$ for events with an input true mass of $M_G$ that satisfy the selection requirements and is referred to as the transfer function (TF).
The TF provides a smooth transformation between the true and reconstructed invariant mass spectrum by modelling the detector effects.
The convolution becomes a product as the true masses $M_{G}$ are discrete and the natural widths can be neglected as the resonances are described by the narrow-width-approximation (NWA) over the range of true masses considered in this analysis.

The cross-sections and branching ratios for the gravitons in the CW/LD signal are taken from Ref.~\cite{Giudice:2017fmj}, where the effects of heavy graviton cascade decays into lighter graviton pairs are included in the total width when calculating the diphoton and dielectron branching ratios.
The additional contributions to the signal from these cascade decays into diphoton and dielectron final states are not considered.
 
The TF and acceptance times efficiency terms are derived following the methodologies of the previous ATLAS dielectron and diphoton resonance searches~\cite{EXOT-2018-08,HIGG-2018-27} and are briefly outlined below.
These terms are derived using the DY MC samples and the resonant single-graviton MC samples for the dielectron and diphoton channels, respectively.
 
For the dielectron channel, the detector response is defined in the TF with respect to the relative dilepton mass resolution ($m_{\ell\ell} - m_{\ell\ell}^{\textrm{true}})/m_{\ell\ell}^{\textrm{true}}$, where $m_{\ell\ell}^{\textrm{true}}$ is the generated dilepton mass at Born level before final-state radiation.
The mass resolution is parameterised as the sum of a Gaussian distribution and a Crystal Ball function~\cite{Oreglia:1980cs,Skwarnicki:1986xj}.
The Gaussian component in this function describes the central peak of the detector response and the Crystal Ball component is used to model the effects of bremsstrahlung in the dielectron system.
The parameterisation of the relative dilepton mass resolution is determined using a simultaneous fit to the resolution function to the DY MC events.
The DY MC sample is separated into 200 bins of equal size in $m_{\ell\ell}^{\textrm{true}}$ based on a logarithmic scale in the dielectron mass range from 130~\GeV to 6000~\GeV.
The mass resolution fit is repeated to evaluate the uncertainty in the fit parameters by shifting individually the lepton energy and momentum scale and resolutions by their uncertainties.
Although the resolution function is derived on spin-1 events, the results are found to be compatible with the true response coming from the spin-2 gravitons.
 
For the diphoton channel, the TF describes the detector resolution of the invariant mass using a double-sided Crystal Ball (DSCB) function~\cite{Oreglia:1980cs, Oreglia:1981fx} with the parameters expressed as a function of the resonance mass, $m_{X}$.
The DSCB parameters are extended to lower masses  relative to Ref.~\cite{HIGG-2018-27} as the CW/LD signals considered in this analysis can contain graviton masses below 500~\GeV, which were not considered in the previous analysis.
 
Analytic templates allow for signal shapes to be created for any choice of $k$ and $M_{5}$.
These templates are confirmed to agree with the MC samples generated at the specific points in the $k$--$M_{5}$ plane discussed in Section~\ref{sec:data}.
Examples of the expected signal shapes in each analysis channel are shown in Figure~\ref{fig:signal_models} for $k=1200$~\GeV, $M_{5} = 3000$~\GeV and for $k=2000$~\GeV, $M_{5} = 8500$~\GeV.
 
\begin{figure}[htb!]
\centering
\subfloat[]{\includegraphics[width=.48\columnwidth]{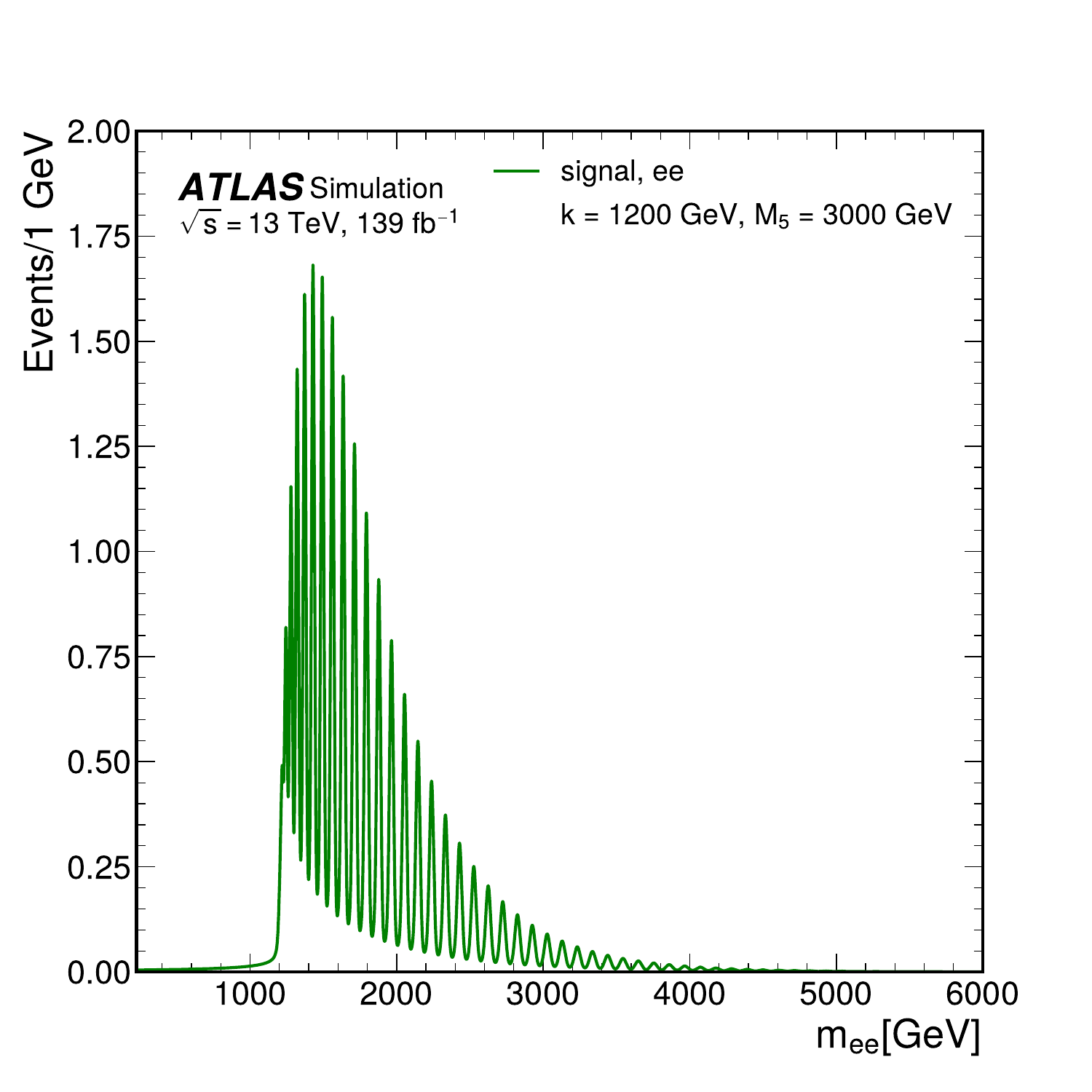}}
\subfloat[]{\includegraphics[width=.48\columnwidth]{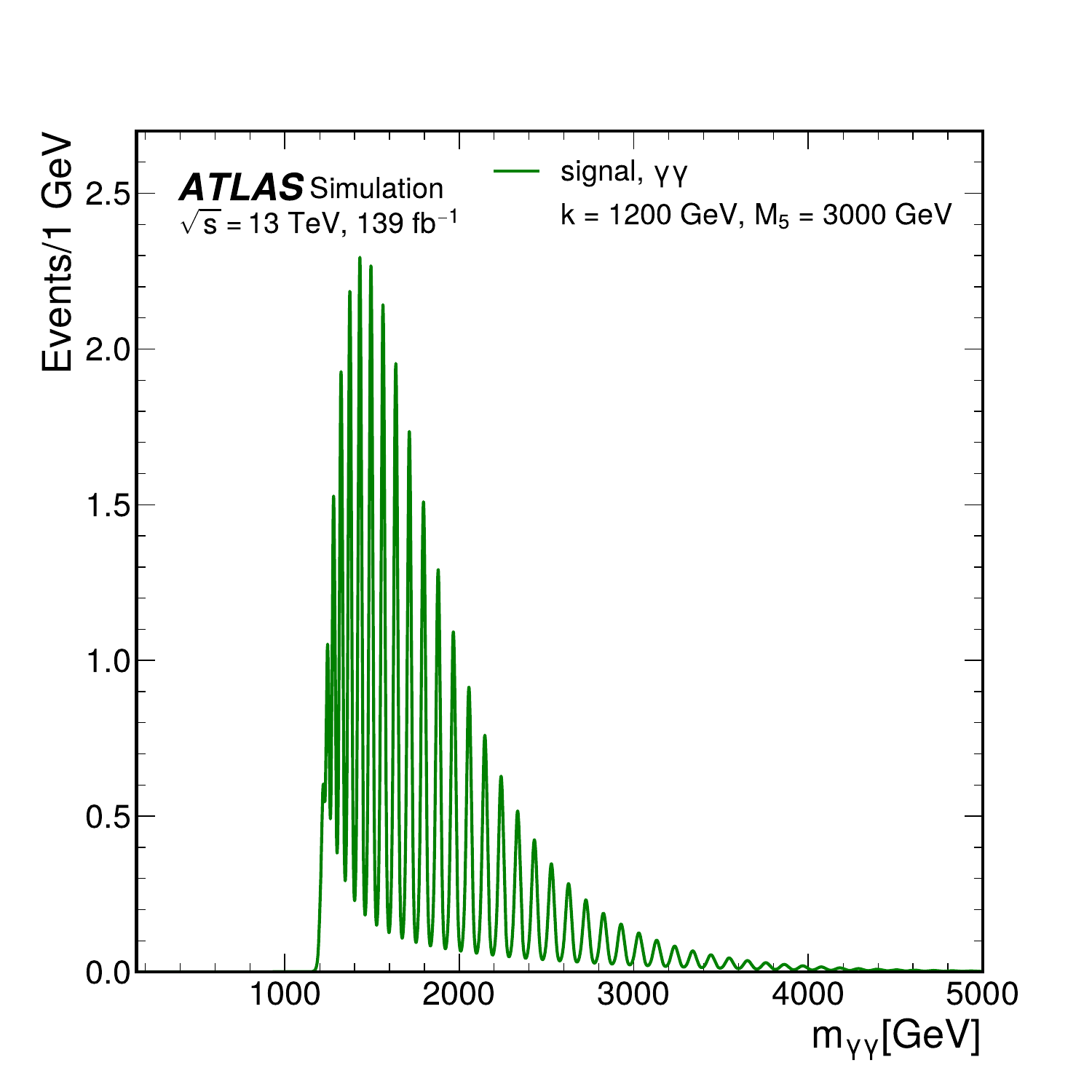}} \\
\subfloat[]{\includegraphics[width=.48\columnwidth]{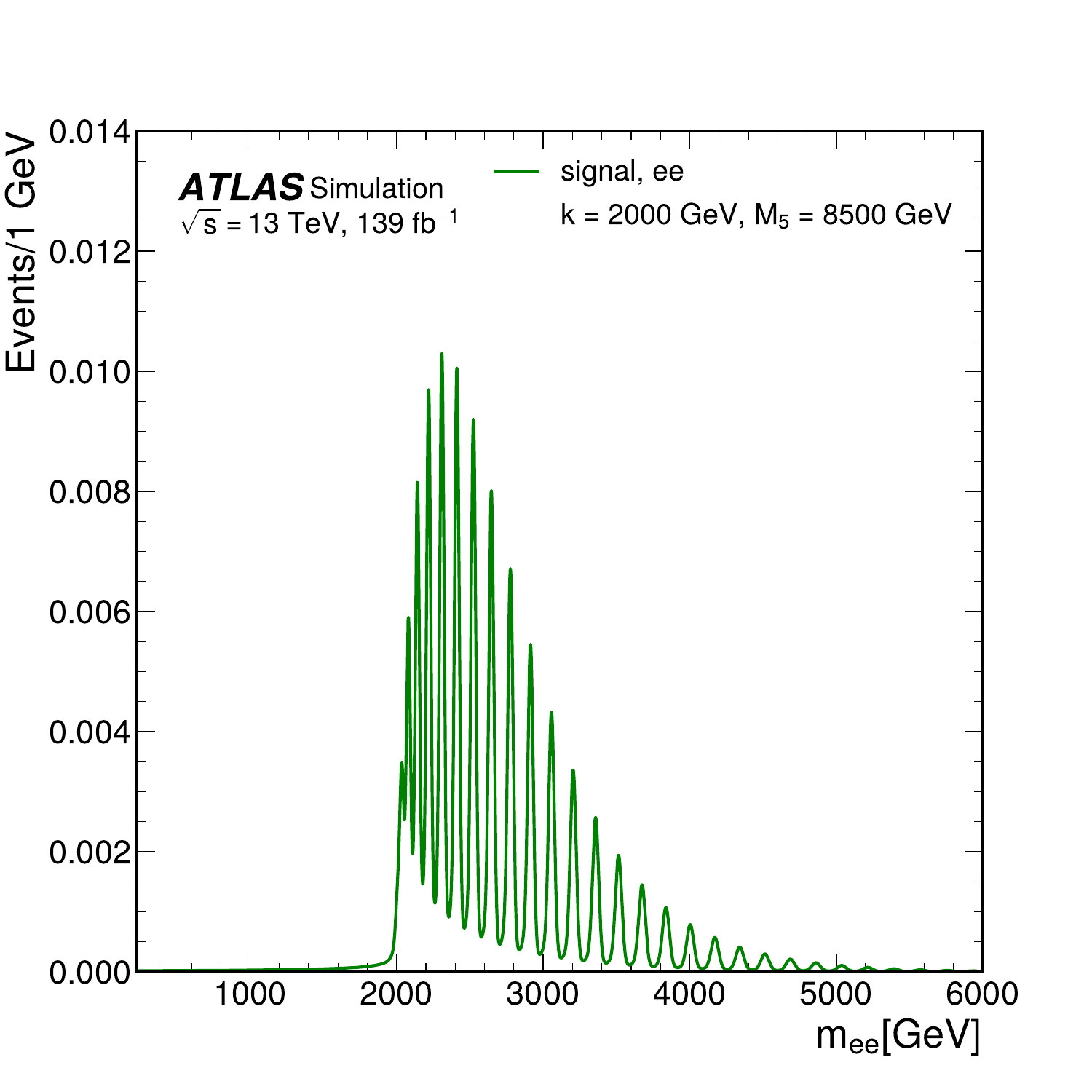}}
\subfloat[]{\includegraphics[width=.48\columnwidth]{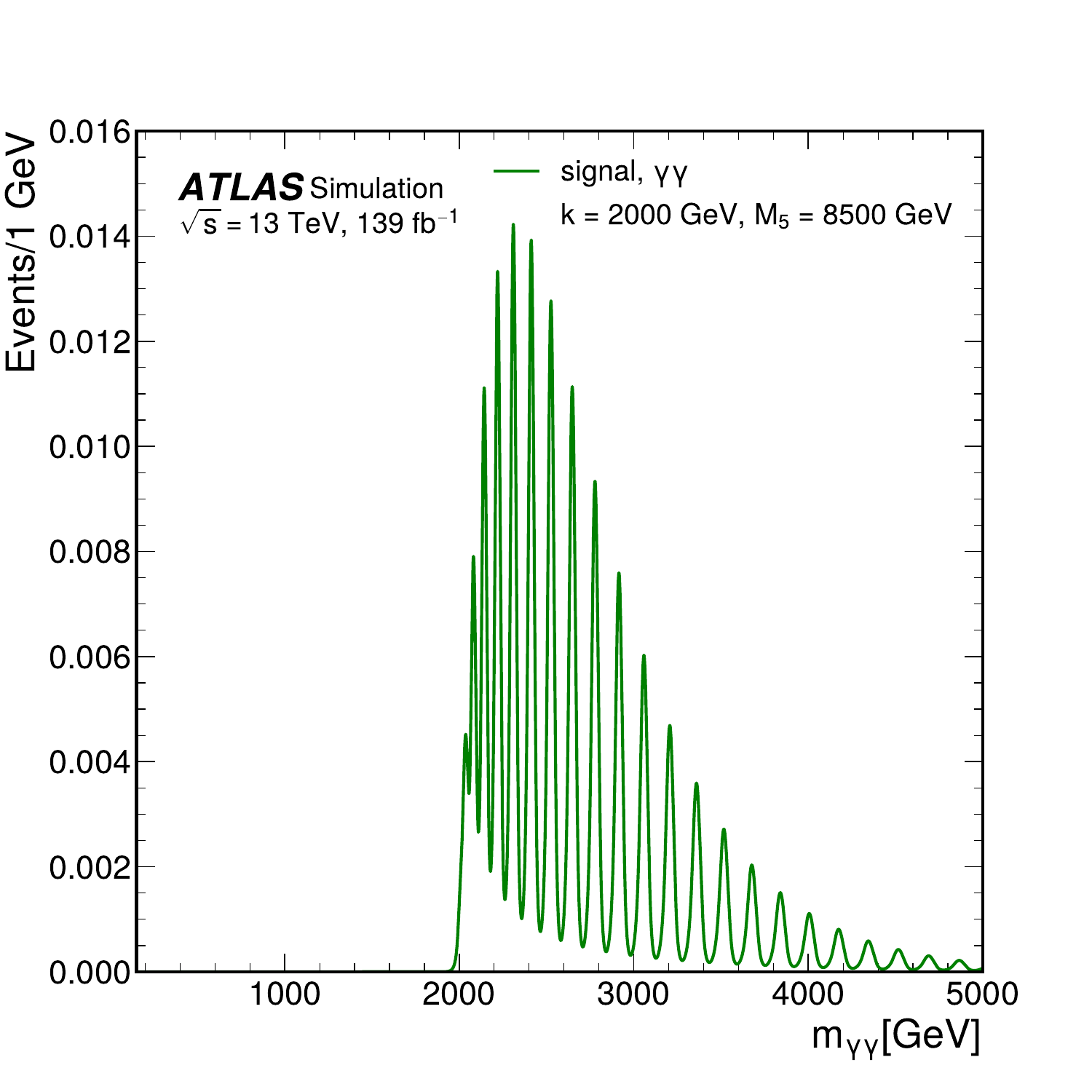}}
\vspace{-0.3cm}
\caption{The expected signal invariant mass shape in the (a) $ee$ and (b) $\gamma\gamma$ channels, where the CW/LD parameter values $k=1200$~\GeV and $M_{5}=3000$~\GeV are used to generate the analytic signal template shapes.
The expected signal invariant mass shape for CW/LD parameter values $k=2000$~\GeV and $M_{5}=8500$~\GeV are shown in the (c) $ee$ and (d) $\gamma\gamma$ channels.
From these two different signal points, it can be seen that the parameter $k$ determines the onset of the KK graviton spectrum.
}
\label{fig:signal_models}
\end{figure}


\section{Background modelling}
\label{sec:backgroundmodeling}

The diphoton and dielectron search channels in this analysis use the same data-driven background estimation technique based on a functional form fit to the data.
Each analysis channel uses a single functional form to model the background across the full invariant mass spectrum.
The analytic form for the background in each channel is determined from a fit to the background expectation derived from MC simulation and control-region data.
 
For the dielectron channel, the functional form is taken from the analysis in Ref.~\cite{EXOT-2018-08}, where the fit was performed on a background template with contributions from the MC estimate of the DY, $t\bar{t}$, diboson, and single-top processes and the data-driven estimate of the reducible backgrounds arising from the multijet and $W$+jet processes.
The function is parameterised as a function of $m_{\ell\ell}$ and is given by:
\begin{eqnarray}
\label{eq:GlobalFit_ee}
f_{ee}(m_\mathrm{\ell\ell}) = f_{\mathrm{BW},Z}(m_\mathrm{\ell\ell}) \cdot \left(1 - x\right)^{b} \cdot x^{\sum_{i=0}^{3} p_i \textrm{log}(x)^{i}}
\end{eqnarray}
where $x = m_{\ell\ell}/\sqrt{s}$ and $f_{\mathrm{BW},Z}(m_{\ell\ell})$ is a non-relativistic Breit--Wigner function with $M_Z = 91.1876\,\GeV$ and $\Gamma_Z = 2.4952\,\GeV$~\cite{ParticleDataGroup:2018ovx}.
The $b$ and $p_{i}$ parameters are fixed to the values from the fit to data in Ref.~\cite{EXOT-2018-08} and the $(1 - x)^{b}$ term ensures that the background fit tends to zero as $x \to 1$, consistent with the expectation from the collision energy of the LHC\@.
The background function is normalised such that its integral corresponds to the total number of events in the signal region.
 
For the diphoton channel, the largest background component arises from the non-resonant production of photon pairs ($\gamma\gamma$ events). Smaller contributions to the total background come from events containing a photon and a jet misidentified as a photon ($\gamma j$ events) and from events with two jets where both of the jets are misidentified as photons ($jj$ events).
The relative contribution from each of these processes is determined using a two-dimensional sideband method described in Ref.~\cite{STDM-2012-05}; the overall $\gamma\gamma$ purity increases as a function of $m_{\gamma\gamma}$ from around 89\% at 150--200 \GeV to around 97\% above 400 \GeV.
The relative uncertainty in the purity ranges from 0.5\%--3\% and the remainder of the background is dominated by $\gamma j$ events.
 
The analytic form for the diphoton channel is chosen from the family of functions used in previous analyses to describe the diphoton invariant mass spectrum~\cite{HIGG-2016-08}, following the procedure described in Ref.~\cite{HIGG-2018-27}.
The background function is described by the following form
\begin{eqnarray}
\label{eqn:powlog1}
f(x;b,a_0,a_1) = (1-x^{1/3})^{b} x^{a_0 + a_1 \textrm{log}(x)}
\end{eqnarray}
where $b$, $a_0$, $a_1$ are free parameters and $x = \mgg/\sqrt{s}$.
 
The dielectron and diphoton invariant mass distributions passing all selections are shown in Figure~\ref{fig:ee_yy_bkg_sig_fit} along with the background-only fits, example clockwork signals for each channel, and the corresponding signal-plus-background fits.
 
\begin{figure}[!htb]
\centering
\subfloat[]{\includegraphics[width=0.48\columnwidth]{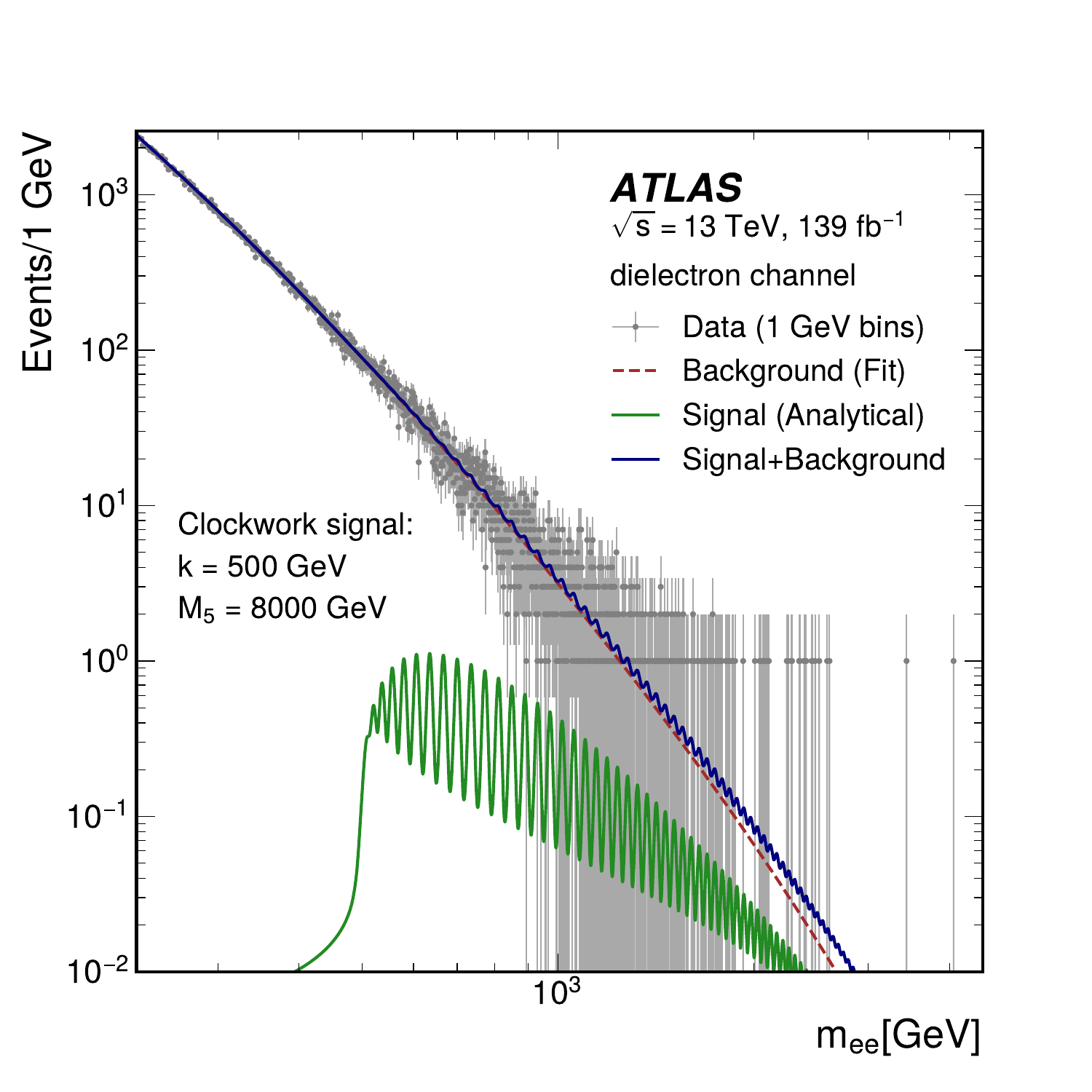}}
\subfloat[]{\includegraphics[width=0.48\columnwidth]{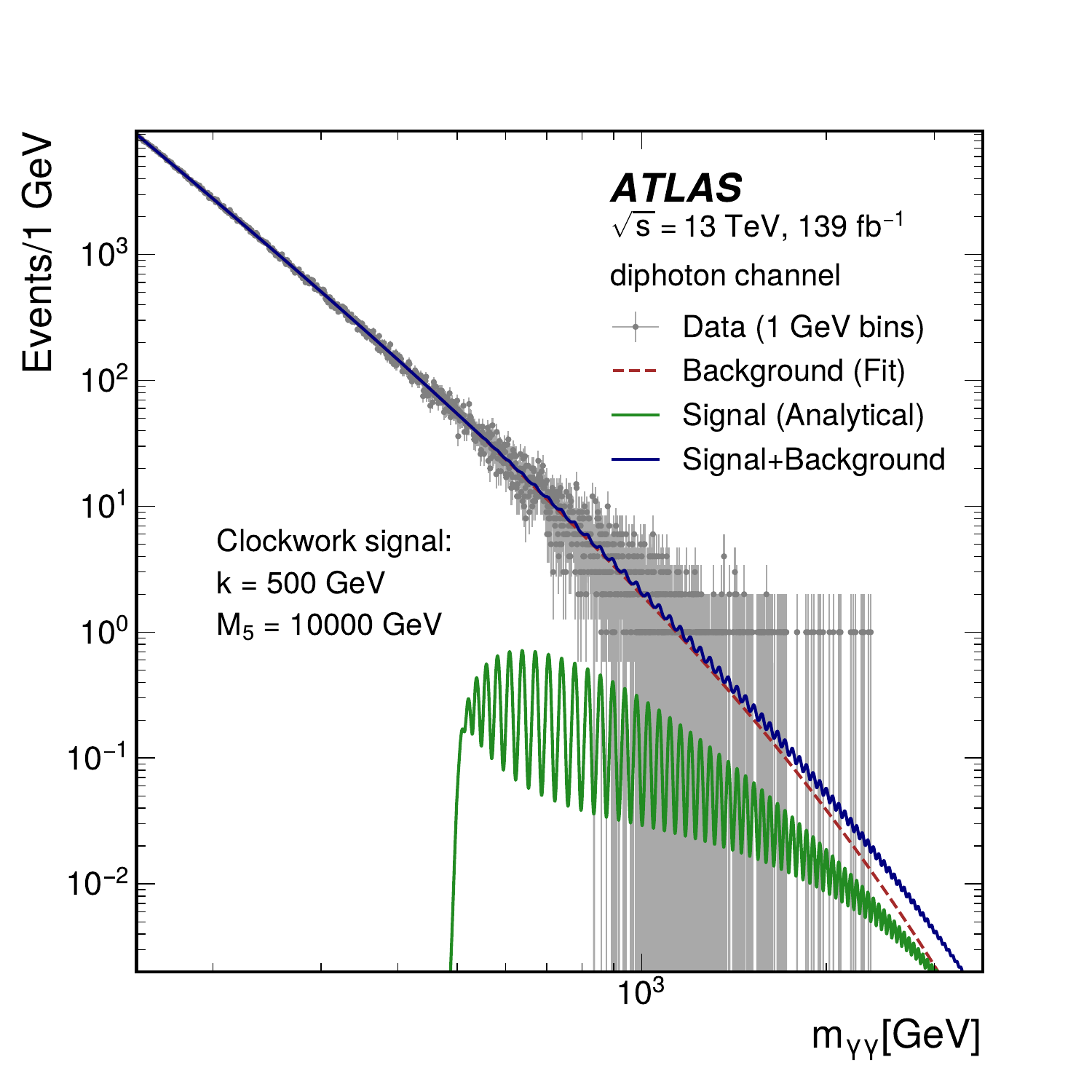}}
\caption{(a) The dielectron and (b) diphoton invariant mass distributions for events passing the full event selection, with the respective background-only fit parameterisation, an analytical clockwork signal, and the signal-plus-background parametrisation.
The clockwork signal parameters are selected close to the sensitivity limit of this analysis.
For the dielectron channel in (a) the clockwork signal is given for $k=500$~\GeV\ and $M_{5}=8000$~\GeV, while the diphoton channel in (b) shows a signal with $k=500$~\GeV\ and $M_{5}=10000$~\GeV.
}
\label{fig:ee_yy_bkg_sig_fit}
\end{figure}


\section{Uncertainties}
\label{sec:uncertainties}

The previous dilepton~\cite{EXOT-2018-08} and diphoton~\cite{HIGG-2018-27} data-driven analyses extracted full MC-based uncertainties, which are used in this search to construct systematically-varied distributions as discussed in Section~\ref{sec:toys}.
The variations considered in both the dielectron and diphoton channels come from theoretical and experimental uncertainties in the simulated backgrounds, uncertainties in the detector response and signal yield, and uncertainties induced by the data-driven background estimates.
 
The following theoretical uncertainties are considered for variations in the DY and $\gamma\gamma$ components: the variations of the nominal PDF set, variations of the factorisation and renormalisation scales, and the effect of choosing different PDF sets.
The variations in the nominal PDF set for the DY background are determined using 90\% CL CT14NNLO PDF error set, while the PDF variations for the $\gamma\gamma$ background are constructed from all of the replicas within the NNPDF3.0 NNLO set.
The factorisation and renormalisation scales are varied independently by a factor of two to estimate the perturbative uncertainties for these background MC samples.
Theoretical uncertainties in the DY component from the variation of the strong coupling, electroweak corrections, and photon-induced corrections~\cite{Martin:2004dh} are also considered.
An additional uncertainty is applied in the dielectron channel to account for the theoretical cross-section uncertainties of the \ttbar process.
 
The experimental uncertainties considered in the diphoton (dielectron) analysis channel include uncertainties on the luminosity determination, pile-up profile, trigger efficiency, photon (electron) energy scale and resolution, and the efficiency of photon (electron) identification and isolation requirements.
More details of the determination of the experimental systematic uncertainties can be found in Refs.~\cite{EXOT-2018-08,HIGG-2018-27}.
 
The systematic uncertainties in the detector response are determined by varying the energy scale and energy resolution of the electrons and photons~\cite{EGAM-2018-01}.
The effect of the experimental uncertainties in the signal yield, based on the variations in the acceptance times efficiency, are also considered.
An additional uncertainty is also derived for the signal based on the internal variations of the PDF set.
 
The largest data-driven uncertainty in the diphoton channel is a shape uncertainty that corresponds to the uncertainty in the relative contribution of the reducible $\gamma j$ background component.
This uncertainty is estimated by varying the measured fraction of the $\gamma\gamma$ background according to its uncertainty from the two-dimensional sideband method mentioned in Section~\ref{sec:backgroundmodeling}.
A data-driven uncertainty in the normalisation of the reducible background component (referred to as the ``fake'' background) in the dielectron channel is also derived, following the methodology of Ref.~\cite{EXOT-2016-05}.
 
In this analysis, each systematic uncertainty is either naturally two-sided or is symmetrised to provide a two-sided systematic uncertainty.
The variations from systematic uncertainties in both analysis channels are smoothed using a sliding window method~\cite{EXOT-2018-08} with a width of 300~\GeV.
The value of the variation at each invariant mass is taken as the average of all points in the range [$m$ $-$ 150~\GeV, $m$ $+$ 150 GeV].
For invariant masses near the edge of the distribution, the input is extended by replicating the value of the variation at the edge.
 
The impact of the systematic uncertainties in the final result is quantified by changes in the test statistic values.
The largest impact on the test statistic values comes from the theoretical uncertainties in the DY and diphoton continuum backgrounds due to the variations of the underlying PDF\@.
The experimental uncertainties are sub-dominant to the theoretical uncertainties across the full mass range considered in each analysis channel.


\section{Generation of pseudo-experiments}
\label{sec:toys}

The analysis uses pseudo-datasets as inputs for the CWT, where one scalogram is generated for each pseudo-dataset.
These scalograms are the basic ingredient for the statistical procedures used in this analysis and are also used for estimating the systematic uncertainties.
The pseudo-datasets (or ``toys'') are generated with background-only models based on the functional forms from Section~\ref{sec:backgroundmodeling} and with signal-plus-background models using the signal models discussed in Section~\ref{sec:sigmod} and the background-only models from Section~\ref{sec:backgroundmodeling}.
These toys are produced with finely-binned (1~\GeV) mass spectra separately for the diphoton and dielectron final states.
For the toys containing a signal, a specific choice of CW/LD model parameters is used.
The generated pseudo-datasets are hereafter called statistical-only toys (stat-toys) or systematic toys (syst-toys) depending on whether systematic variations are considered when constructing the toy.
 
To make the stat-toys, the analytic shapes are constructed containing either a background-only shape or a signal-plus-background shape.
The background component is given a normalisation corresponding to the data normalisation in the signal region of each analysis channel.
Next, a Poisson-fluctuated toy distribution is generated from the analytic shape chosen.
Each of these toys is then treated as a proxy for the real data in the subsequent statistical analysis discussed in Section~\ref{sec:statanalysis}.

To make the syst-toys, a similar procedure is followed as with the stat-toys, but starting from an alternative description of the background-only and signal-plus-background shapes.
This analysis follows the procedure introduced in Ref.~\cite{EXOT-2019-16}, which is described briefly below.
 
To make one syst-toy for the background-only case, the alternative background shape (called hereafter an uncertainty template) is built from the systematic variations that arise from the sources of both experimental and theoretical uncertainties as discussed in Section~\ref{sec:uncertainties}.
In the background-only case, these systematic variations are given in the invariant mass space as shape variations around the background description from simulation, as obtained in Refs.~\cite{EXOT-2018-08,HIGG-2018-27}.
A relative variation is defined as the relative difference between the up or down variation and the background description from simulation.
To obtain the absolute systematic difference for  the nominal background description from data defined in Sections~\ref{sec:sigmod} and~\ref{sec:backgroundmodeling}, the relative variation is multiplied by the nominal background description from data.
Each uncertainty template is constructed from the nominal background description from data, summed with the weighted absolute systematic differences.
The weights are randomly sampled from a Gaussian distribution, with a mean of zero and standard deviation of one.
The number of such weights in each uncertainty template is equal to the number of systematic variations considered.
 
To replicate the procedures given in Refs.~\cite{EXOT-2018-08, HIGG-2018-27}, the resulting uncertainty template is Poisson-fluctuated to generate one uncertainty pseudo-dataset.
This uncertainty pseudo-dataset is then fit with the background model from either Eq.~(\ref{eq:GlobalFit_ee}) or Eq.~(\ref{eqn:powlog1}), depending on the analysis channel.
This procedure results in one smooth alternative background description, replacing the fits done in Refs.~\cite{EXOT-2018-08, HIGG-2018-27}.
This alternative background description is then used to generate one respective background-only uncertainty pseudo-dataset.
 
To make one syst-toy in the signal-plus-background case, a similar procedure is followed with minor modifications.
The procedure starts from the signal systematic variations discussed in Section~\ref{sec:uncertainties}.
The nominal signal shape is described by Eq.~(\ref{eq:grav_signal_modeling}).
The absolute systematic difference for each variation is calculated directly for the nominal signal shape.
The signal uncertainty template is constructed from the nominal signal shape, summed with the Gaussian-weighted absolute systematic differences.
Each signal-plus-background uncertainty template is constructed as the sum of one smooth alternative background description, as discussed above, and one signal uncertainty template.
The result is Poisson-fluctuated to generate one respective signal-plus-background uncertainty pseudo-dataset.
 
These procedures are repeated to generate background-only and signal-plus-background syst-toys ensembles, where each toy in these ensembles reflects a different (random) combination of all uncertainties.
The syst-toys ensembles are used in the subsequent statistical analysis discussed in Section~\ref{sec:statanalysis}.


\section{Continuous wavelet transforms}
\label{sSec:CWT}

Continuous wavelet transforms are used in this search to convert the diphoton and dielectron invariant mass spectra into mass versus frequency information in the form of scalograms before applying the machine learning techniques discussed in Section~\ref{sec:statanalysis}.
An overview of the CWT method used in this search is detailed below.
 
The CWT is a measure of similarity between a chosen wavelet and a signal.
The wavelet is used to scan the signal for different frequencies of the wavelet throughout the invariant mass spectrum.
 
The wavelet $\psi(x)$ is a basis function localised in both the mass and frequency space and the CWT of a signal $f(m)$ at a scale\footnote{The scale $\alpha$ is inversely proportional to frequency, where higher scales represent the low frequency component and vice versa.} $\alpha$ and translational parameter $\beta$ is given by a projection on $\psi(x)$ for different $\alpha$ and $\beta$ values~\cite{Beauchesne:2019tpx}:
\begin{equation}
\label{eq:CWTdef}
W(\alpha, \beta) = \frac{1}{\sqrt{\alpha}}\int_{-\infty}^{+\infty}f(m)\, \psi^*\left(\frac{m - \beta}{\alpha}\right)dm \,
\end{equation}
A scalogram can be produced by taking the norm of the coefficient $W(\alpha, \beta)$ for all values of $\alpha$ and $\beta$.
Therefore, the CWT in this analysis defines how much of a certain frequency is in the signal at a given invariant mass bin.

The \emph{Morlet wavelet}~\cite{Morlet} is used in this analysis as the choice of $\psi(x)$ in \cref{eq:CWTdef}.
The Morlet wavelet is chosen because it is Gaussian-shaped in the frequency domain, which minimises possible edge effects that could be interpreted as signal, as opposed to wavelets with sharp boundaries that can introduce edge effects~\cite{COHEN201981}. It consists of a localised wave packet and is given by:
\begin{equation}
\psi(x) \equiv \frac{1}{\sqrt{B\pi}}\, e^{-x^2/B} \left(e^{i 2\pi C x} - e^{-\pi^2 B C^2}\right),
\end{equation}
where $B$ and $C$ are constants which are chosen to be 2 and 1 respectively.
This choice of $C$ ensures that the CWT of a signal reaches a maximum when the signal wavelength approximately equals the scale $\alpha$.
The Morlet Wavelet is shown in Figure~\ref{Fig:morlet}.
 
\begin{figure}[htb!]
\centering
\includegraphics[width=.68\columnwidth]{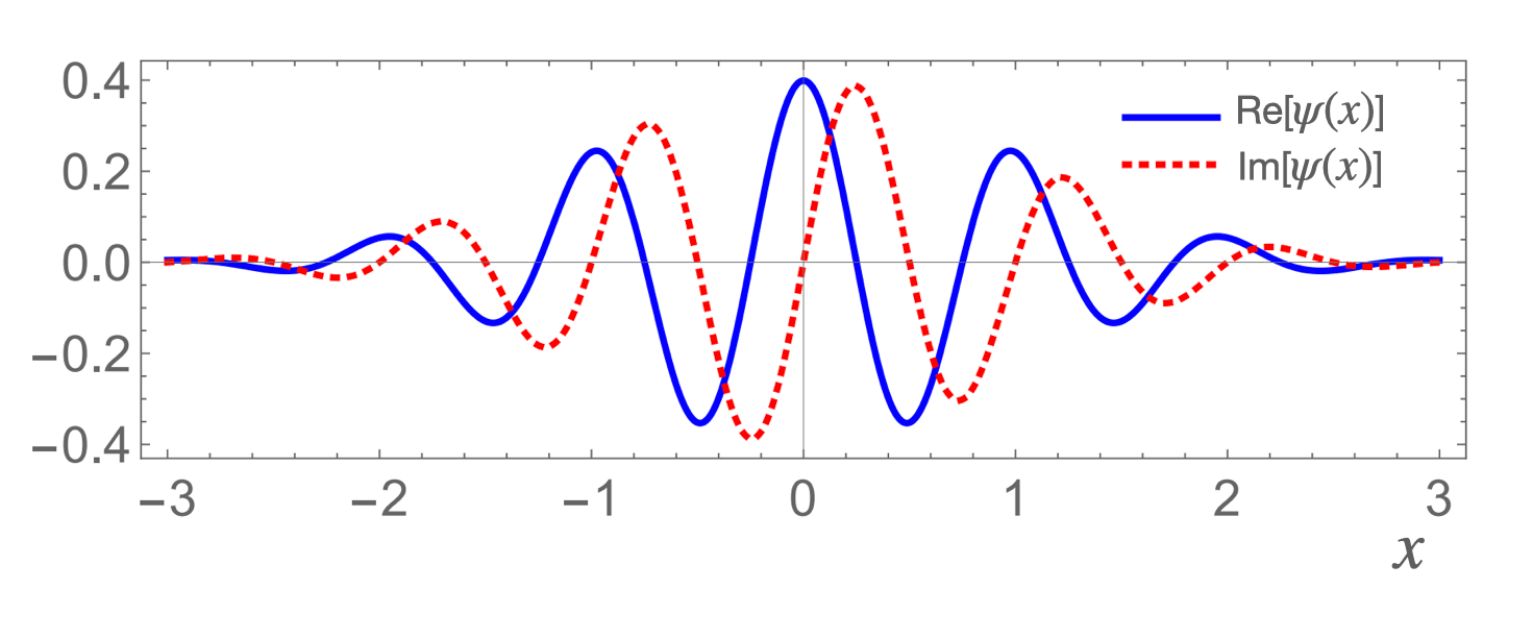}
\caption{An illustration of the complex Morlet Wavelet used in the continuous wavelet transform in this analysis, split into real and imaginary components.}
\label{Fig:morlet}
\end{figure}

Example scalograms for the dielectron and diphoton channels can be seen in Figures~\ref{Fig:eeyySigBkgNoFluc} (without statistical fluctuation) and~\ref{Fig:eeyySigBkg} (with statistical fluctuation).
These scalograms are created after applying the CWT on a background-only distribution (Figures~\ref{Fig:eeyySigBkgNoFluc_ee_bkgonly},~\ref{Fig:eeyySigBkgNoFluc_yy_bkgonly},~\ref{Fig:eeyySigBkg_ee_bkgonly} and~\ref{Fig:eeyySigBkg_yy_bkgonly})
and on a signal-plus-background distribution (Figures~\ref{Fig:eeyySigBkgNoFluc_ee_bkgsig},~\ref{Fig:eeyySigBkgNoFluc_yy_bkgsig},~\ref{Fig:eeyySigBkg_ee_bkgsig} and~\ref{Fig:eeyySigBkg_yy_bkgsig}).
The saturated area in the high-scale region occurs because the background distribution is not flat.
Therefore, the background may still appear as periodic for very large scales, that is, those scales with small frequencies.
The sharp transition seen between low $|W(\alpha, \beta)|$ and high $|W(\alpha, \beta)|$ values occurs because the $|W(\alpha, \beta)|$ values change quickly relative to the scale chosen for the $z$-axis.
 
The signal contribution in \Cref{Fig:eeyySigBkg_ee_bkgsig,Fig:eeyySigBkg_yy_bkgsig} is clearly seen as a local small ``island,'' discernible from the continuum of the background shown in \Cref{Fig:eeyySigBkg_ee_bkgonly,Fig:eeyySigBkg_yy_bkgonly}, even when realistic statistical fluctuations are included.
This signal-island represents the locality of the signal in both the mass and the scale (or frequency) spaces.
Fixing $M_5$ and increasing the value of $k$ shifts the signal-island in two ways.
The signal-island is shifted horizontally to higher masses because the signal turn-on point in mass is roughly equal to the value of $k$.
The signal-island is also shifted vertically towards higher scales because the spacing in the KK tower increases with $k$.
Fixing $k$ and changing $M_5$ only determines the prominence of the island, that is, it becomes more distinct with decreasing $M_5$ because of the inverse relationship between $M_{5}$ and the signal cross-section.
Changing the mass resolution due to detector effects (within the uncertainties) only affects the $|W(\alpha, \beta)|$ values of the signal-island and this change is effectively equivalent to changing $M_5$.

\begin{figure}[htb!]
\centering
\subfloat[]{\includegraphics[width=.48\columnwidth]{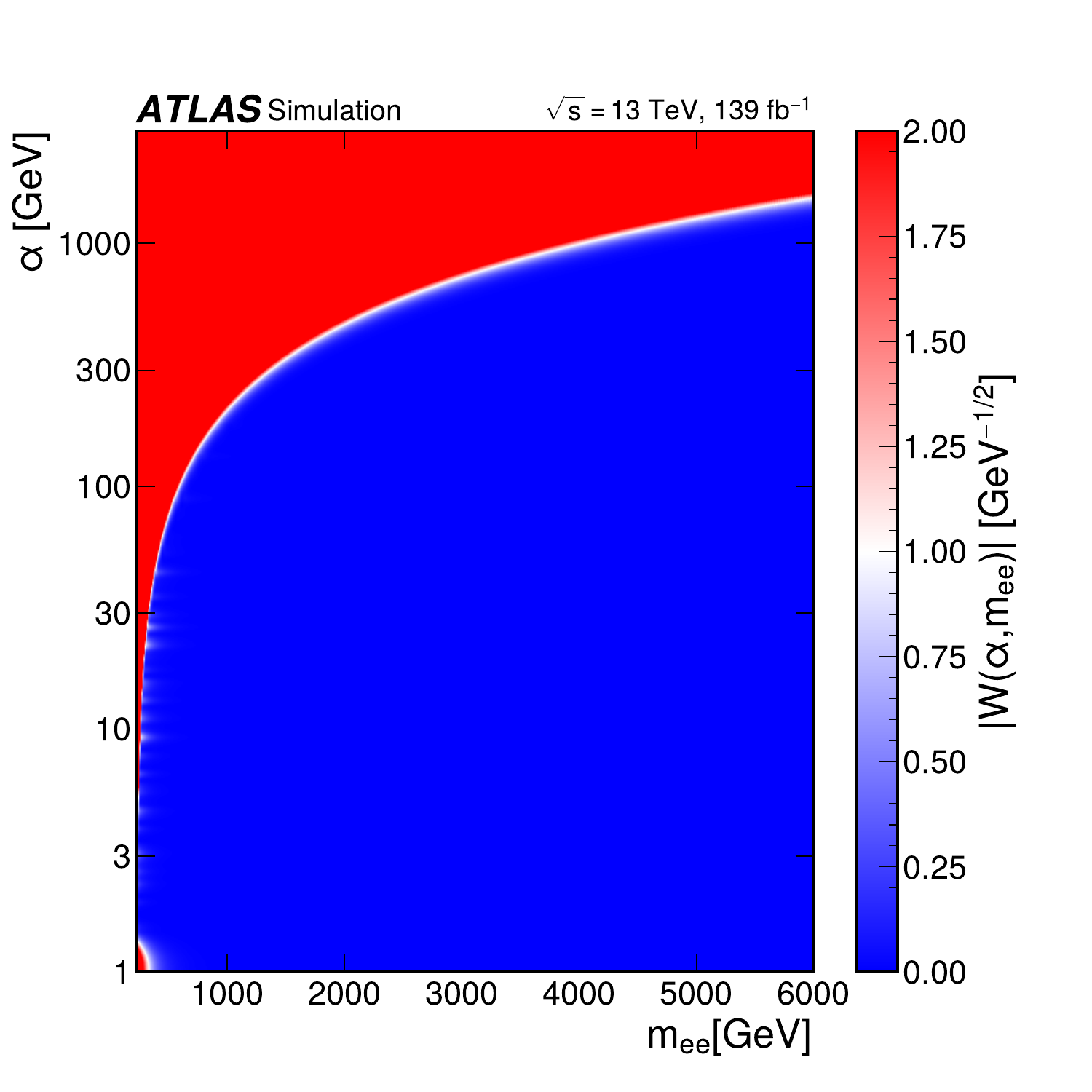}\label{Fig:eeyySigBkgNoFluc_ee_bkgonly}}
\subfloat[]{\includegraphics[width=.48\columnwidth]{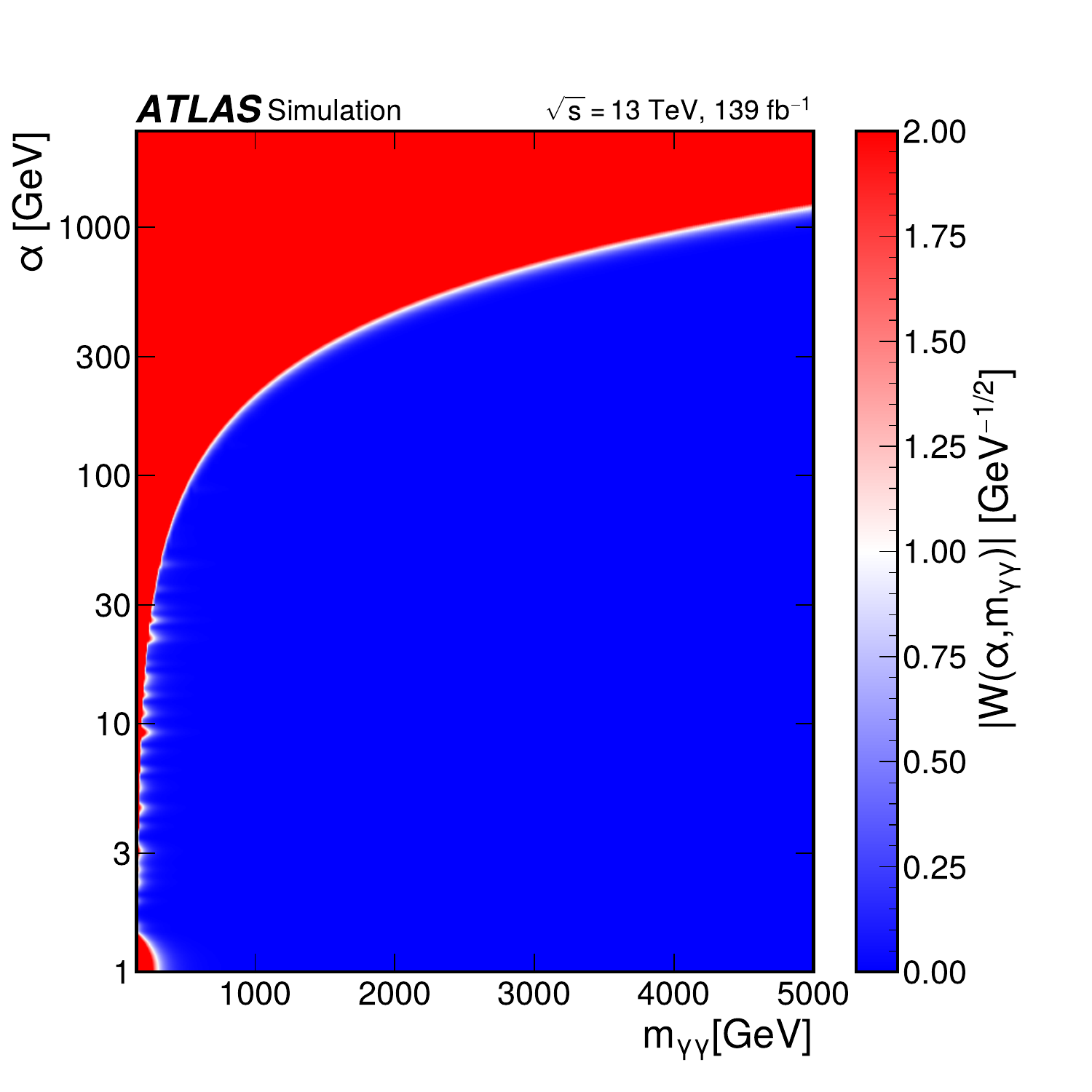}\label{Fig:eeyySigBkgNoFluc_yy_bkgonly}}\\
\vspace{-0.3cm}
\subfloat[]{\includegraphics[width=.48\columnwidth]{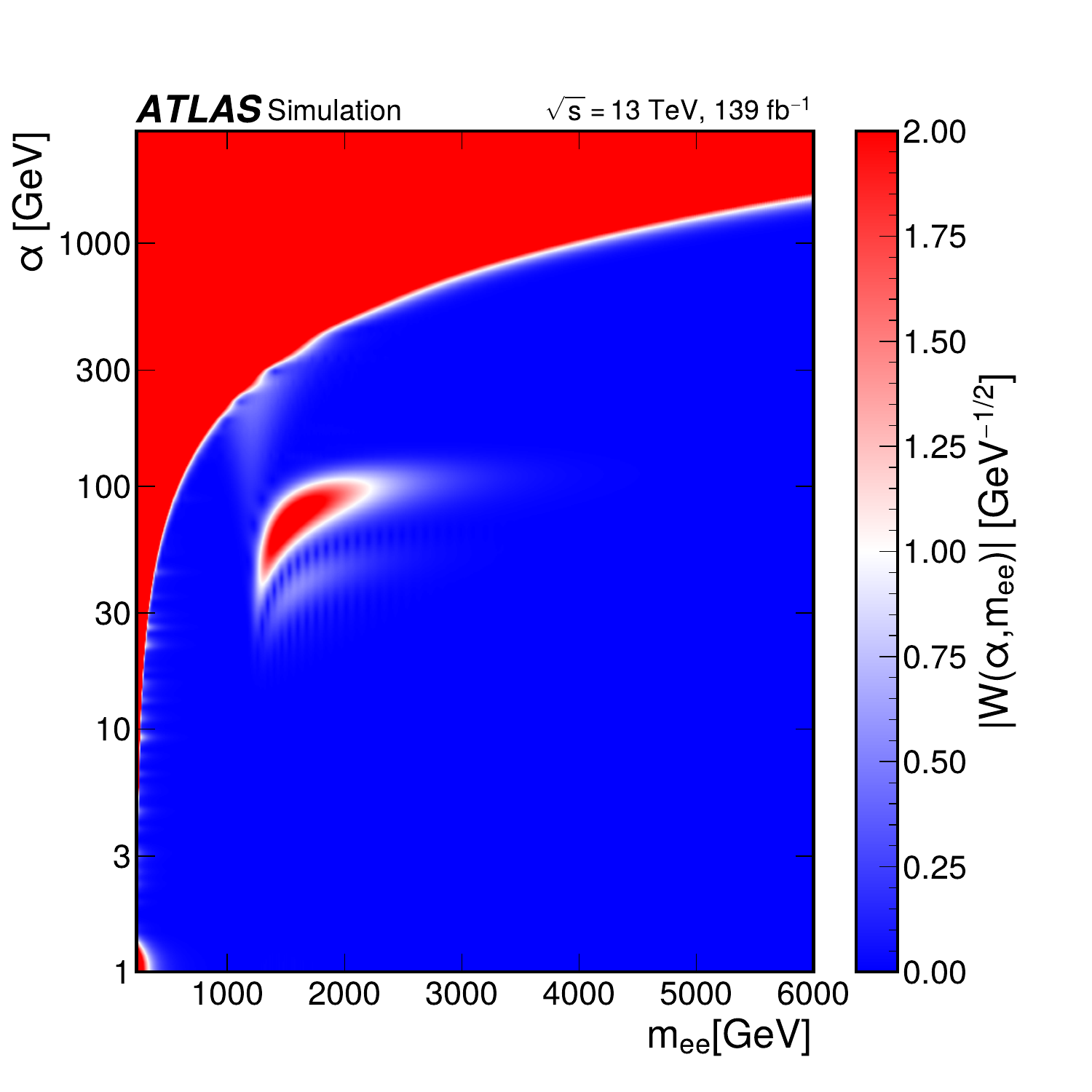}\label{Fig:eeyySigBkgNoFluc_ee_bkgsig}}
\subfloat[]{\includegraphics[width=.48\columnwidth]{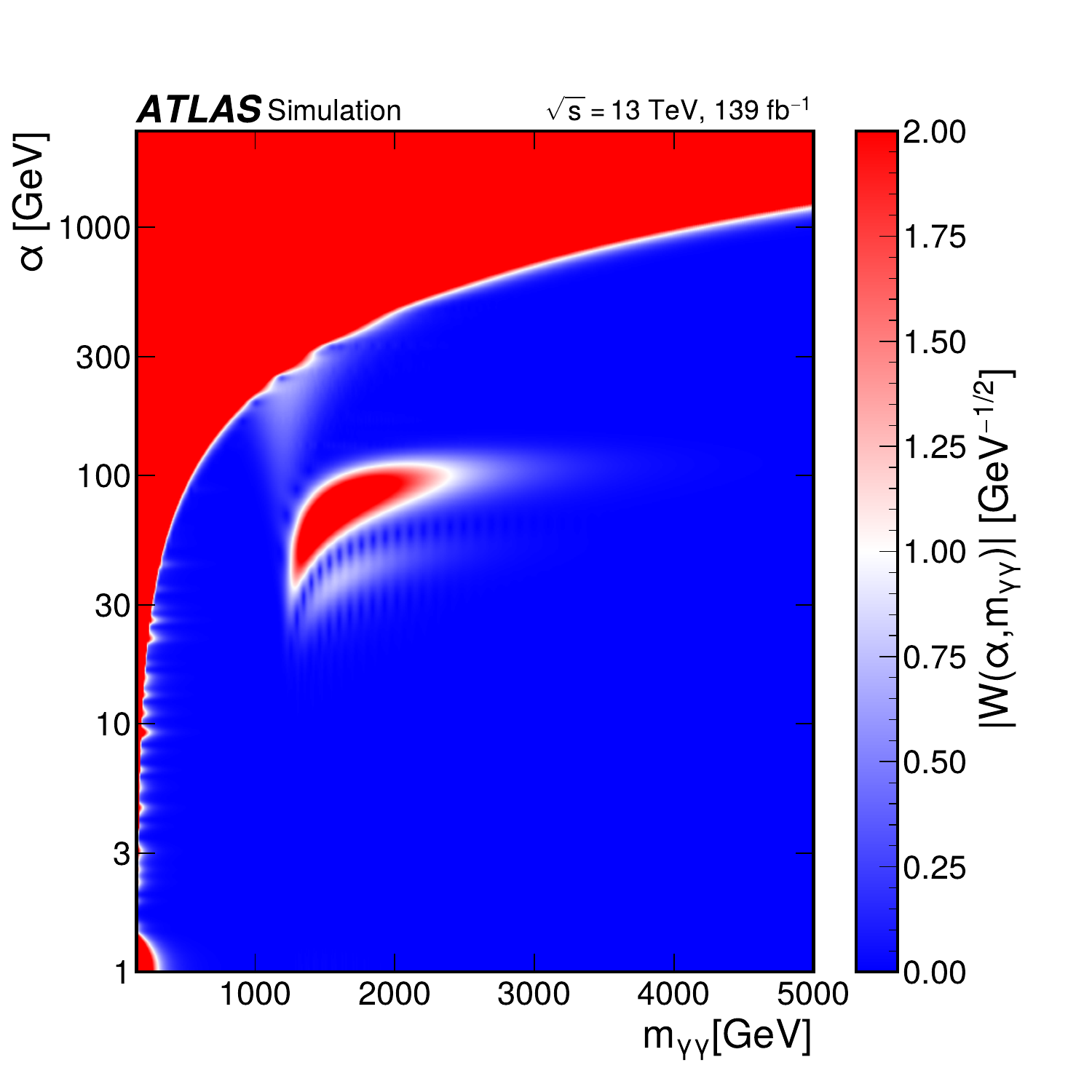}\label{Fig:eeyySigBkgNoFluc_yy_bkgsig}}
\caption{Scalogram output of the CWT of (a) dielectron and (b) diphoton background-only toy experiments and (c) dielectron and (d) diphoton signal-plus-background toy experiments with $k=1200$~\GeV and $M_5 = 3000$~\GeV. These toys are produced without Poisson fluctuations.
The signal contribution in (c) and (d) manifests as a localised ``island'' in mass and scale, discernible from the continuum of the background shown in (a) and (b).
The scalograms are provided here with a mass binning of 1~\GeV.
Here $\alpha$ is the CWT scale parameter and $W(\alpha,\beta)$ are the wavelet coefficients defined in Eq.~(\ref{eq:CWTdef}), where the invariant mass of each channel takes the role of $\beta$.}
\label{Fig:eeyySigBkgNoFluc}
\end{figure}

\begin{figure}[htb!]
\centering
\subfloat[]{\includegraphics[width=.48\columnwidth]{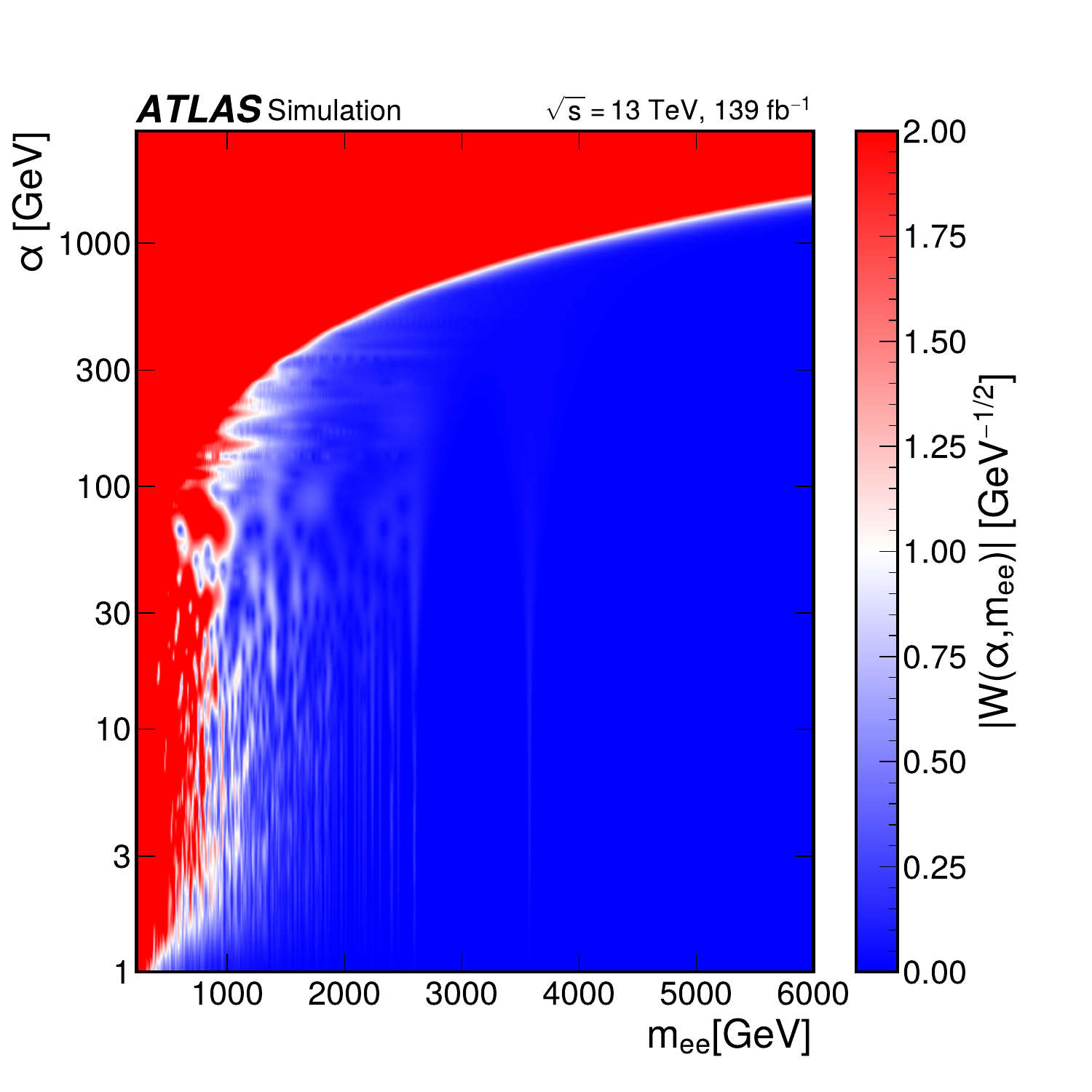}\label{Fig:eeyySigBkg_ee_bkgonly}}
\subfloat[]{\includegraphics[width=.48\columnwidth]{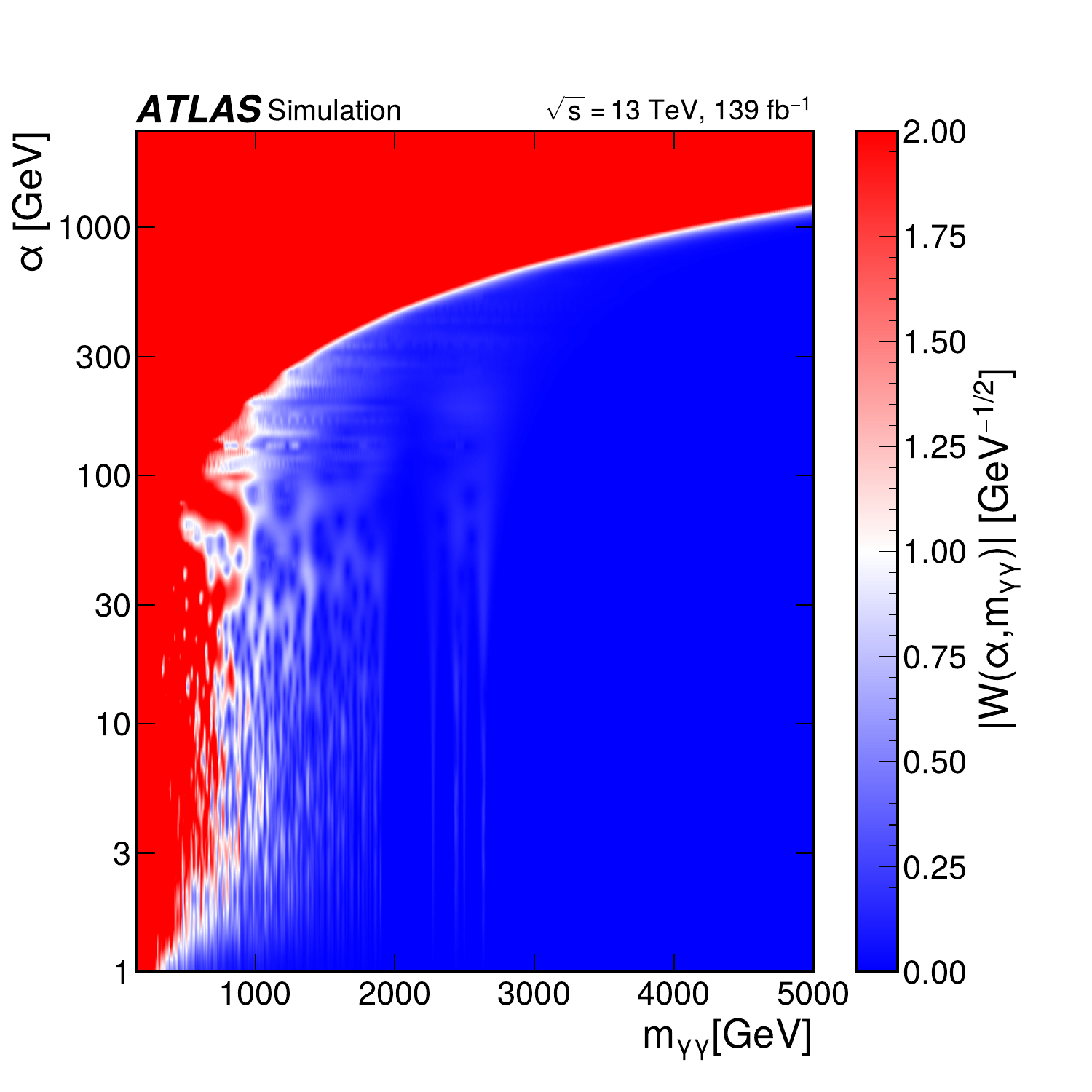}\label{Fig:eeyySigBkg_yy_bkgonly}}\\
\vspace{-0.3cm}
\subfloat[]{\includegraphics[width=.48\columnwidth]{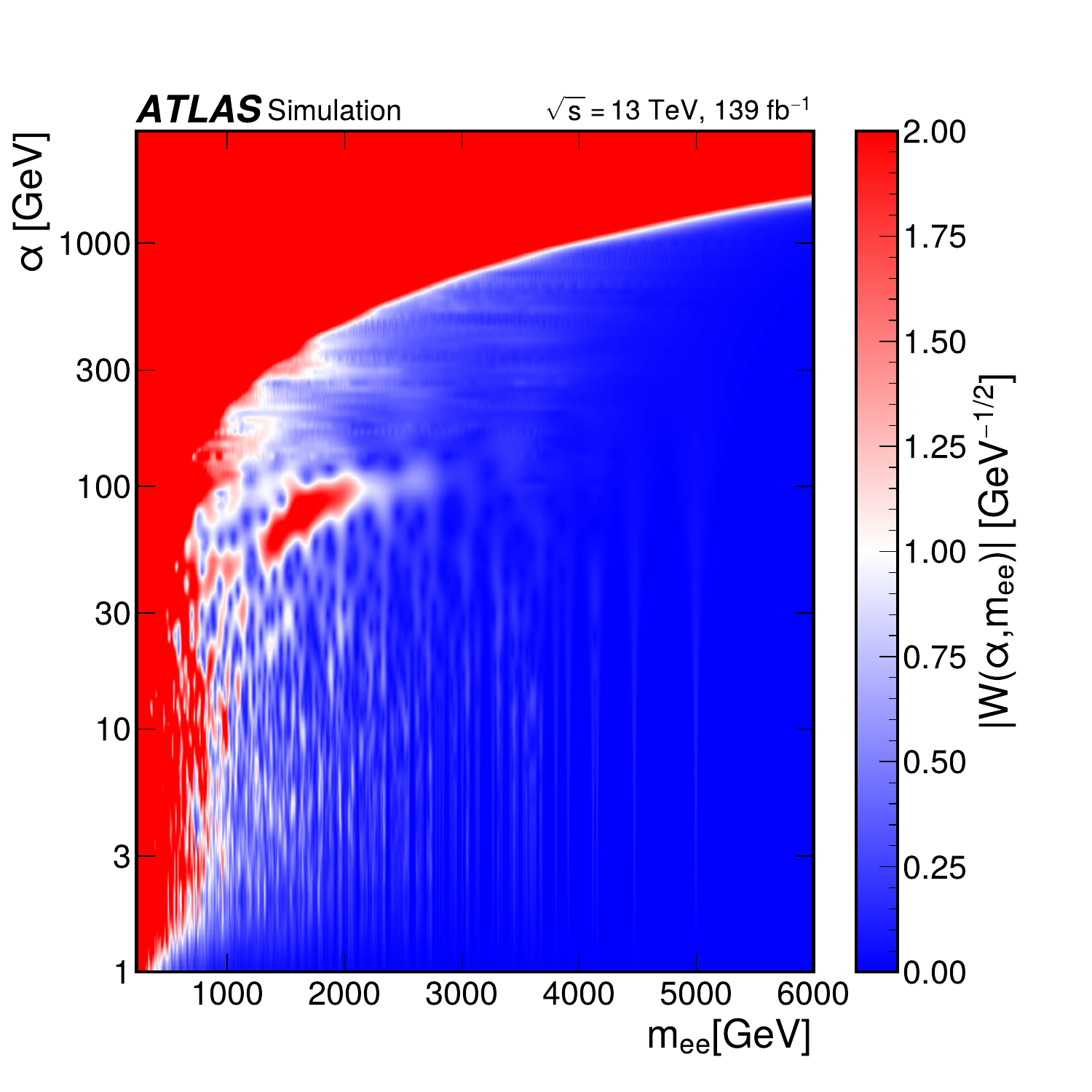}\label{Fig:eeyySigBkg_ee_bkgsig}}
\subfloat[]{\includegraphics[width=.48\columnwidth]{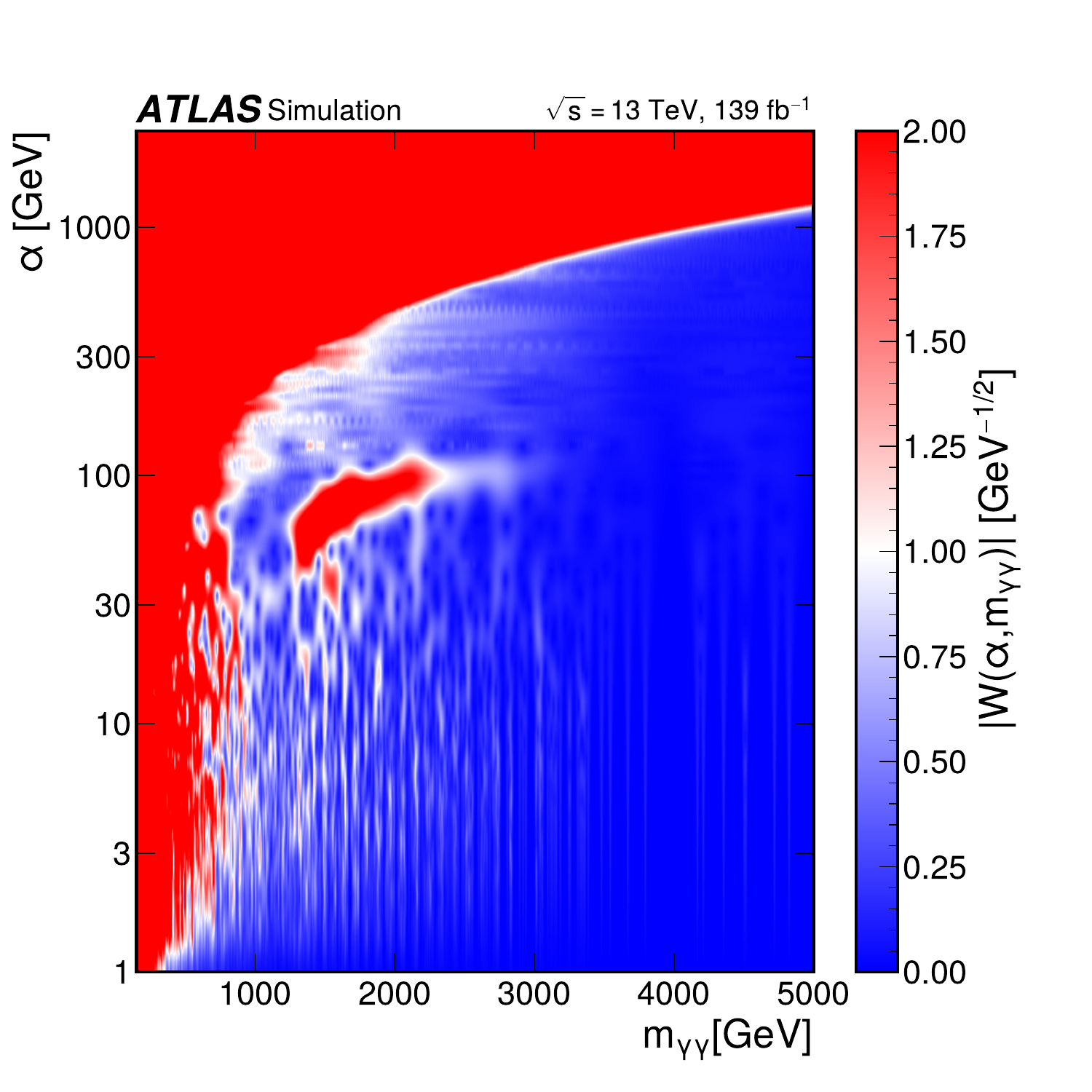}\label{Fig:eeyySigBkg_yy_bkgsig}}
\caption{Scalogram output of the CWT of (a) dielectron and (b) diphoton background-only toy experiments and (c) dielectron and (d) diphoton signal-plus-background toy experiments with $k=1200$~\GeV and $M_5 = 3000$~\GeV.
These toys are produced with Poisson fluctuations arising from the statistical uncertainty of the data sample.
These fluctuations are shown to dilute the presence of a signal.
The signal contribution in (c) and (d) manifests as a localised ``island'' in mass and scale, discernible from the continuum of the background shown in (a) and (b).
The scalograms are provided here with a mass binning of 1 GeV.
Here $\alpha$ is the CWT scale parameter and $W(\alpha,\beta)$ are the wavelet coefficients defined in Eq.~(\ref{eq:CWTdef}), where the invariant mass of each channel takes the role of $\beta$.}
\label{Fig:eeyySigBkg}
\end{figure}


\section{Data analysis}
\label{sec:statanalysis}

The diphoton and dielectron invariant mass distributions from the toys detailed in Section~\ref{sec:toys} are transformed into images using the CWT method described in \cref{sSec:CWT} before being used to train a neural network (NN).
The diphoton and dielectron images are treated as independent channels and are trained separately.
Two types of convolutional NNs are used in this search, a classifier NN and an autoencoder (AE), which are discussed in \cref{sSec:Classifier} and \cref{sSec:Autoencoders}, respectively.
In both of the cases, the training is done using stat-toys, while the prediction is performed using syst-toys to evaluate the impact of the systematic uncertainties in the NN responses.
As discussed in Section~\ref{sec:toys}, the syst-toys already include a proper statistical representation of the data.
 
The classifier NN is used to probe for the periodic signals of the CW/LD model specifically (denoted here as model-dependent).
On the other hand, the AE NN is used to search for more general anomalies in the data (denoted here as model-independent).
Both of the neural networks used in this analysis are based on the setups given in Ref.~\cite{Beauchesne:2019tpx}.
The neural networks are implemented in Keras~\cite{chollet2015keras} using the TensorFlow backend~\cite{Abadi:2016kic}.
The ADAM~\cite{Kingma:2014vow} optimiser is used to minimise the loss functions in each NN setup.
The search is performed using a general form of a test statistic derived from the NN output itself.
 
The loss functions associated with each of the NNs play the role of the likelihood function in a typical analysis.
The loss function used in the classifier NN is the Binary Cross-Entropy (BCE)~\cite{Cohen:2017exh}, while the AE NN uses the Mean Squared Error (MSE) loss function.
The MSE loss function is further discussed in Section~\ref{sSec:Autoencoders}.
The training and validation loss are recorded as a function of the number of epochs and are inspected to verify that each NN gives good agreement in the output of the loss function when comparing the training and validation datasets.

In this analysis, the NN output in the classifier setup and the value of the loss function in the AE setup are used as test statistics.
To set the exclusion limits on the model-dependent parameter space, the modified frequentist method, commonly known as the $\mbox{CL}_s$ method~\cite{Read:2002hq,Junk:1999kv, ParticleDataGroup:2020ssz}, is used.
This choice results in more conservative limits than those obtained by the standard $p$-value procedure~\cite{Read:2002hq,Junk:1999kv}.
The dielectron and diphoton channels are analysed separately due to potential overlaps in their event selections.

\subsection{Classifier NN}\label{sSec:Classifier}
 
One possible way to discover a specific signal in a scalogram is with a classifier NN\@.
In this analysis, the convolutional NN classifier setup from Ref.~\cite{Beauchesne:2019tpx} is implemented for the model-dependent search.
The CWT scalogram of the mass spectrum for each stat-toy, syst-toy, and the signal region data is produced and rebinned to ensure the inputs to the NN remain manageable.
The mass for the NN inputs is rebinned into 40~\GeV wide bins, while the scale is binned logarithmically.
The stat-toy scalograms for the background-only and signal-plus-background cases are used for training the binary classifier convolutional NN, where the output of the network is between $0$ (background-like) and $1$ (signal-plus-background like).

In addition to the signal periodicity of the CW/LD model, these models can introduce a non-resonant tail above the SM background at high dielectron or diphoton invariant masses.
For these high-mass ranges, the periodicity may no longer be as apparent due to the widening of the resonances with mass.
This signal widening occurs because the experimental mass resolution worsens, which causes the merging of the signal mass peaks.
This analysis focuses on the periodic features of the signal by adjusting the procedure such that the subsequent inference does not consider the effects of the non-resonant tail.
 
To achieve this adjustment, a cutoff is applied to the dielectron or diphoton mass distribution to remove events above a certain point in mass prior to training the classifier.
This mass threshold is determined by calculating a signal's local significance, $S/\sigma_B$, where $S$ is the signal yield and $\sigma_B$ is the local per-bin uncertainty.
This uncertainty includes the statistical uncertainty and the systematic uncertainties described in Section~\ref{sec:uncertainties},  added in quadrature, of the background.
The bin in the invariant mass distribution on a local signal peak where the per-bin-significance falls below 50\% of its maximal value is chosen as the threshold.
While the cutoff is independent of $M_5$ by construction, it does rise monotonically with $k$.
Therefore, the cutoff values are calculated for a few points in $k$ and are then fit with a second-order polynomial to allow for interpolation between the $k$ points.
As the long-range signal shape falls rapidly after reaching its maximum, alternative choices close to the nominal choice of 50\% of the maximum local significance have little impact on the result.
A lower significance choice, for example $ S/\sigma_{B} \approx 10\%$ of the maximal value, effectively introduces no cutoff and the subsequent inference remains affected by the presence of the non-resonant tail feature.
A higher significance choice, e.g., $ S/\sigma_{B} \approx 90\%$ of the maximum value, removes the non-resonant tail effectively but also removes a large mass range where the periodicity is still apparent and reduces the search sensitivity to the intended feature.
For the dielectron channel, the 50\% threshold mass value ranges from 1400 to 6000~\GeV for $k$ in the range of 200~\GeV to 5850~\GeV.
For the diphoton channel, the mass threshold ranges from 2000 to 5000~\GeV for $k$ in the range of 150~\GeV to 4900~\GeV.
 
For completeness, the classifier search without mass thresholds is also performed.
In this case, the non-resonant feature of the signal may be present, while the periodicity may be unresolvable in the high mass region.
The CWT and NN procedure can still provide some discrimination power between the signal-plus-background and background-only distributions due to the non-resonant feature and additional oscillations available to describe the signal.
Therefore, the sensitivity for low-$k$ ($k\lesssim1000$~\GeV) signals is expected to be higher than for the case with mass thresholds.
For high-$k$ values, ($k\gtrsim1000$~\GeV), the sensitivities of the two methods are expected to be almost identical.

\subsection{Autoencoder NN}\label{sSec:Autoencoders}
A model-independent approach searching for anomalies in scalograms is performed using AEs.
This technique has previously been applied to jet images in Refs.~\cite{Heimel:2018mkt,Farina:2018fyg} and was implemented and suggested for this type of search in Ref.~\cite{Beauchesne:2019tpx}.
The AE compresses a scalogram to a smaller set of parameters, which are then used to reconstruct the input scalogram.
Instead of using the original background-only scalograms, they are standardised such that each bin in the scalogram is remapped to a local p-value in the range 0 to 1.
The remapping per bin is done using the distribution of the original values in the specific bin from multiple scalograms.
Further, to regularise the output, the negative logarithm of the remapped bin value (local p-value) is used to build a new two-dimensional input.
This procedure is identical to the procedure in Ref.~\cite{Beauchesne:2019tpx} and ensures that the loss function is not dominated by the region with high statistical precision.
The AE is trained on background-only, stat-toy remapped scalograms to reproduce the original remapped scalograms ($\bvec{y}$) by minimising the MSE loss function~\cite{Farina:2018fyg}, $L_\text{MSE}$, as defined below:
\begin{equation}
\label{eq:meansquarederror}
L_\text{MSE}(\bvec{y},\bvec{{\hat{y}}}) = \frac{1}{n} \sum_{i=1}^n |\mathrm{y}_i - {\hat{\mathrm{y}}_i}|^2 \ ,
\end{equation}
where $n$ is the number of bins used in the scalogram and $\bvec{{\hat{y}}}$ is the output remapped scalogram from the AE.
 
After the AE model is trained, the AE should be able to approximately reproduce the original remapped scalogram if applied to a typical background-only syst-toy scalogram.
Conversely, the AE may fail to reproduce the scalogram if applied to a data scalogram that contains a signal.
This metric, $L_\text{MSE}$, is used as a test statistic.
In this analysis, the AE is particularly sensitive to signals with periodic structures.

The $L_\text{MSE}$ is also calculated per scalogram bin.
For each bin, an ensemble of the $L_\text{MSE}$ values for the background predictions is generated.
From these ensembles, a local $p$-value is calculated for every scalogram bin.
The statistically significant bins are then visible with $p$-values close to 0.
 
As discussed in Section~\ref{sSec:Classifier}, to focus the search on the periodicity features and avoid possible inference that is based only on non-resonant signals, two sets of selections are introduced.
Unlike in Section~\ref{sSec:Classifier}, these selections cannot be based on a specific signal and should be characterised only based on information pertaining to the SM backgrounds.
In the first method, the non-resonant features are removed by using the statistical error at the tail of the background template shape, $R = 1/\sqrt{B(m_\text{min})}$, where $B(m_\text{min})$ is the integral of the background template shape above some minimum invariant mass $m_\text{min}$.
Different $m_\text{min}$ thresholds are derived for different $R$ values, where for each, the tail of the distribution above the threshold is removed before calculating the CWT\@.
The thresholds and the respective invariant mass ranges for each channel are found in Table~\ref{table:mass_thresholds_AE}.
 
\begin{table}[htb]
\caption{Different precision-driven thresholds considered for each analysis channel and the corresponding mass ranges probed in the model-independent analysis.
For the no threshold scenario, the upper limit in mass is chosen such that the range goes 300~\GeV beyond the last observed data point in the dielectron and diphoton mass spectra.
}
\begin{center}
\begin{tabular}{ccc}
\toprule
Threshold & Dielectron Mass Range [GeV] & Diphoton Mass Range [GeV] \\
\midrule
$ R < 10\%$ & 225--1520 & 150--1380 \\
$ R < 50\%$ & 225--2700 & 150--2400 \\
No threshold & 225--4400 & 150--2700 \\
\bottomrule
\end{tabular}
\end{center}
\label{table:mass_thresholds_AE}
\end{table}
 
This procedure also prevents large statistical fluctuations in the high-mass tail of the background distribution from affecting the $L_\text{MSE}$ of the AE predictions.
 
In the second approach, the scale range of the scalograms is instead modified and the mass range remains unrestricted.
This approach, referred to as scale thresholding, avoids the large-scale region of the scalogram that may include continuum contributions from a signal.
In this method, the bin content in the scalogram is replaced with zero if the scale $\alpha$ in that bin satisfies $\alpha > m/4$, where $m$ denotes the mass value of that bin.
In a similar way, to exclude the scales that are short and cannot be resolved experimentally, the bin content in the scalogram is replaced with zero if the bin satisfies $\alpha < m/100$.
The results from both methods are given in Section~\ref{sec:result}.
 
Possible deviations in the data, arising from SM or beyond the SM effects, from the smoothly falling background model assumed in each analysis channel could be captured in the AE analysis.
However, the sensitivity to any new signal is lessened as this approach is designed to be most sensitive for deviations which appear periodic.


\section{Results}
\label{sec:result}

The scalograms for the data events passing all selections for the diphoton and dielectron channels are shown in Figure~\ref{fig:finalScalograms}.
 
\begin{figure}[h]
\centering
\subfloat[]{\includegraphics[width=0.49\textwidth]{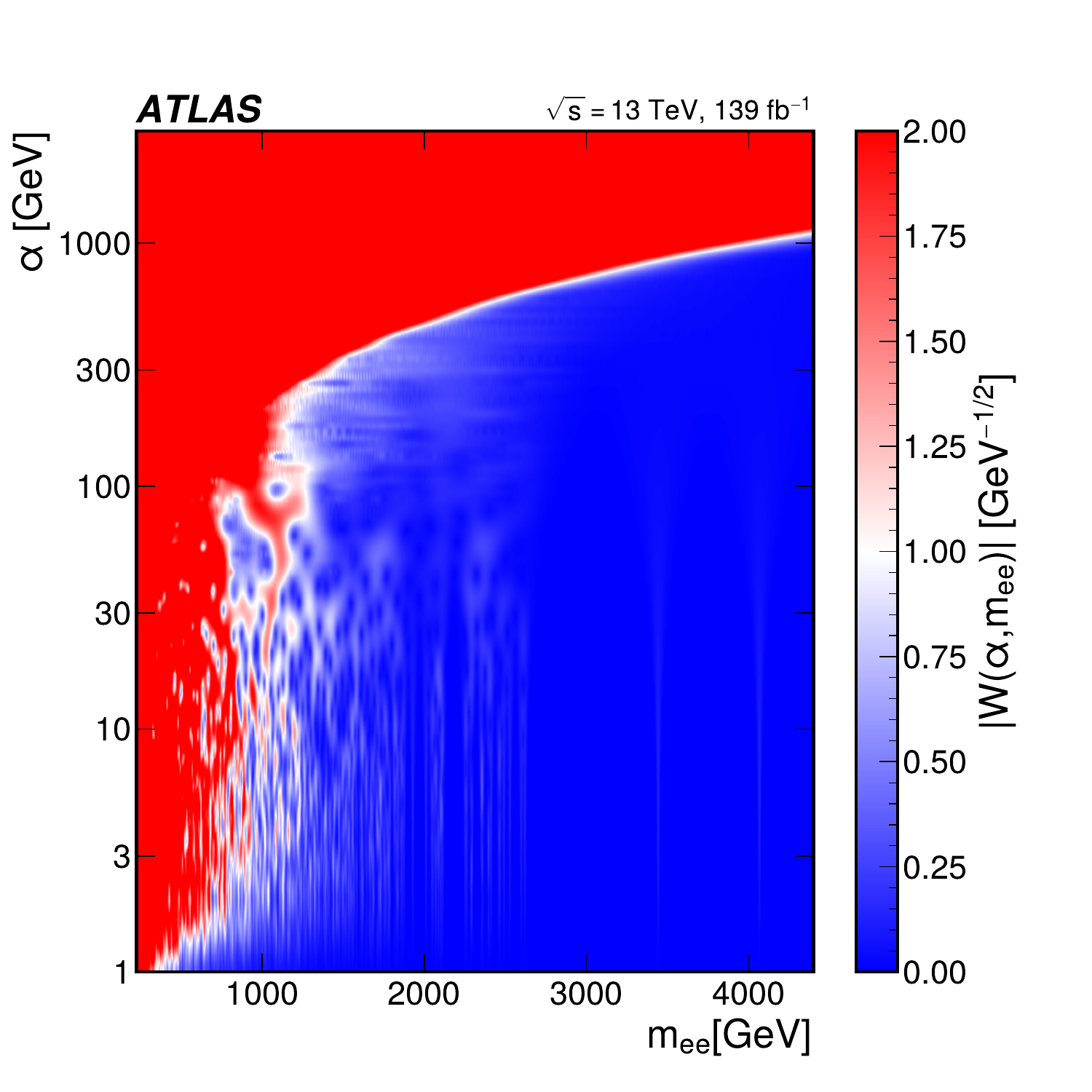}}
\subfloat[]{\includegraphics[width=0.49\textwidth]{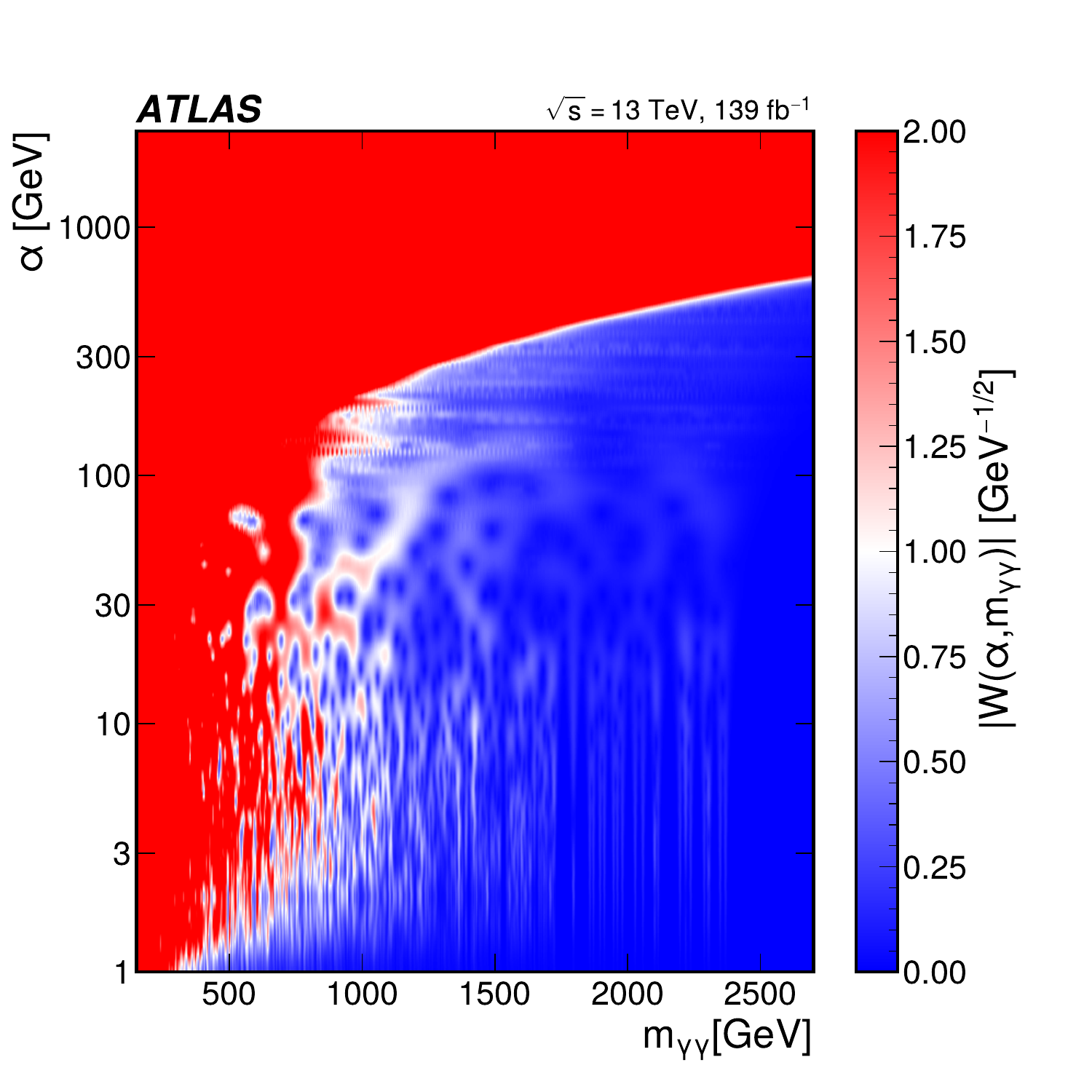}}
\caption{The scalogram output of the CWT for the (a) dielectron and (b) diphoton search channels for the observed signal region data. The scalograms are shown here with a mass binning of 1~\GeV and are rebinned to a coarser binning before running the neural networks.
Here $\alpha$ is the CWT scale parameter and $W(\alpha,\beta)$ are the wavelet coefficients defined in Eq.~(\ref{eq:CWTdef}), where the invariant mass of each channel takes the role of $\beta$.
The range of $m_{ee}$ and $m_{\gamma\gamma}$ is chosen such that the plots cover approximately $300$~\GeV after the last observed data point.
}
\label{fig:finalScalograms}
\end{figure}

The model-independent results from the AE-based search with $R < 50\%$ are presented as the negative logarithm of the local $p$-values corresponding to these data scalograms and are shown in Figure~\ref{fig:localsignificance}.
 
\begin{figure}[htb!]
\centering
\subfloat[]{\includegraphics[width=0.49\textwidth]{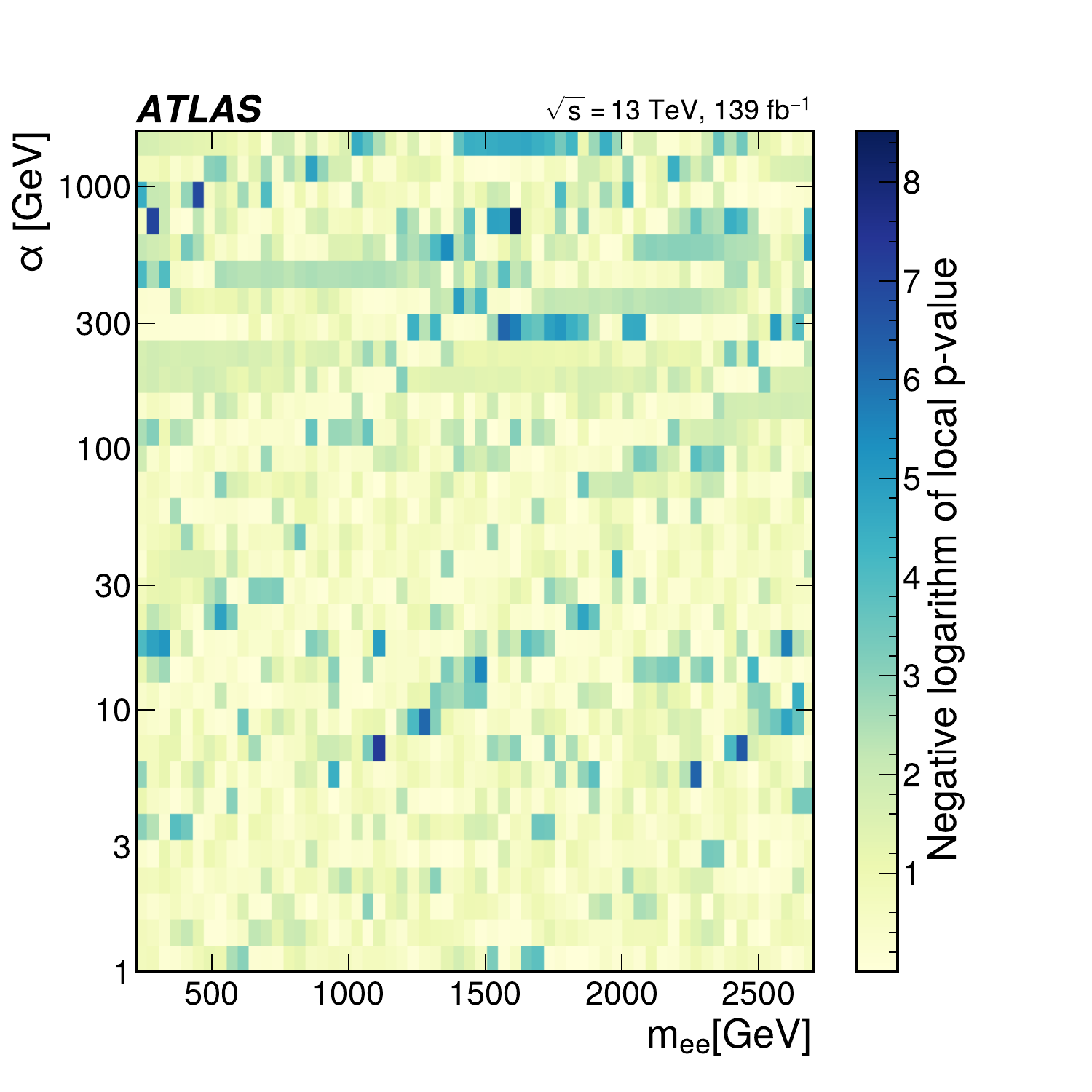}}
\subfloat[]{\includegraphics[width=0.49\textwidth]{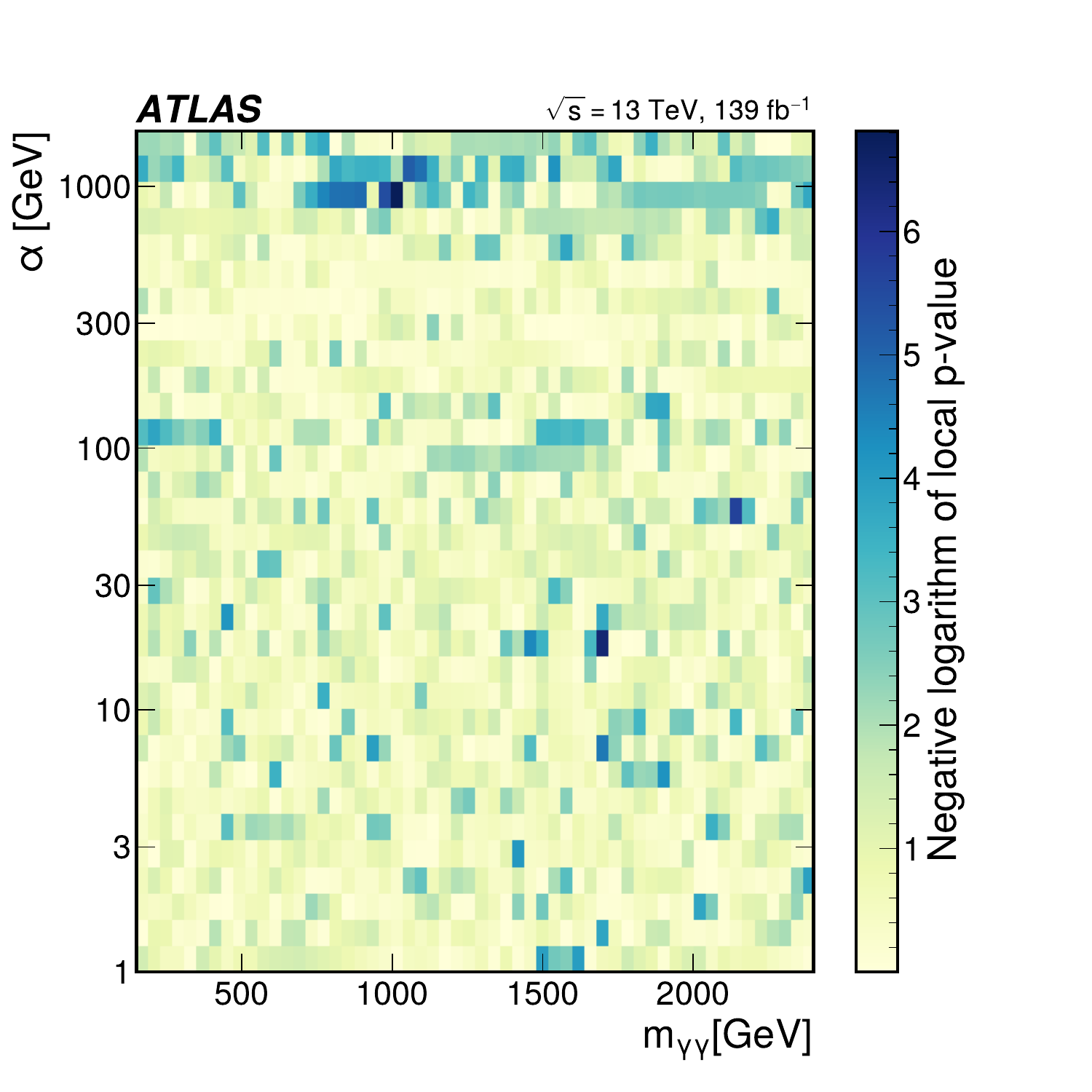}} \\
\caption{The negative logarithm of the local $p$-values calculated for the observed data mass distributions in the (a) dielectron and (b) diphoton search channels for the $R < 50\%$ scenario. 
Here $\alpha$ is the CWT scale parameter.
}
\label{fig:localsignificance}
\end{figure}
 
To obtain the significance for the model-independent results, the $L_\text{MSE}$ is calculated for each original and predicted scalogram pair, with the sum in Eq.~(\ref{eq:meansquarederror}) running over all of the scalogram bins.
An ensemble of $L_\text{MSE}$ values for the background prediction is then generated from many pairs of original and predicted scalograms.
From this ensemble, the significance is calculated for the data scalogram.
 
The significances for the model-independent results in each analysis channel are given in Table~\ref{table:significance_observed}.
No significant deviations from the background-only hypothesis are observed.
The largest deviation from the background-only (SM) hypothesis is found in the dielectron channel with the 50\% threshold and the scale threshold, with an excess corresponding to a significance of $1.5\sigma$.
 
For the AE-based model-independent result, an alternative approach is also tested where the training is performed using syst-toys rather than stat-toys.
In this approach, all steps of the AE-based search after the training match the procedure described in Section~\ref{sSec:Autoencoders}.
The significance values in this alternative approach are found to be comparable to the results given in Table~\ref{table:significance_observed}.
 
\begin{table}[htb!]
\caption{The observed significance in each analysis channel for the different thresholds considered in the model-independent analysis.
Each significance is signed with positive (negative) values indicating the loss from data is higher (lower) than the median loss from the expected background.
}
\begin{center}
\begin{tabular}{ccc}
\toprule
Threshold & Dielectron Significance & Diphoton Significance \\
\midrule
$ R < 10\%$ & 0.4 & --1.8 \\
$ R < 50\%$ & 1.5 & --0.2 \\
No threshold & 0.7 & --0.7 \\
Scale threshold  & 1.5 & --0.6 \\
\bottomrule
\end{tabular}
\end{center}
\label{table:significance_observed}
\end{table}
 
The model-dependent results are also presented for each analysis channel.
No significant deviation from the background-only hypothesis is seen in either of the analysis channels.
Test statistic distributions assuming both a signal model, with $k=1033$~\GeV and $M_{5} = 9000$~\GeV, and the Standard Model are shown in Figure~\ref{fig:teststatdata} and they are compared with the test statistic from the observed data.
\begin{figure}[htb!]
\centering
\includegraphics[width=0.49\textwidth]{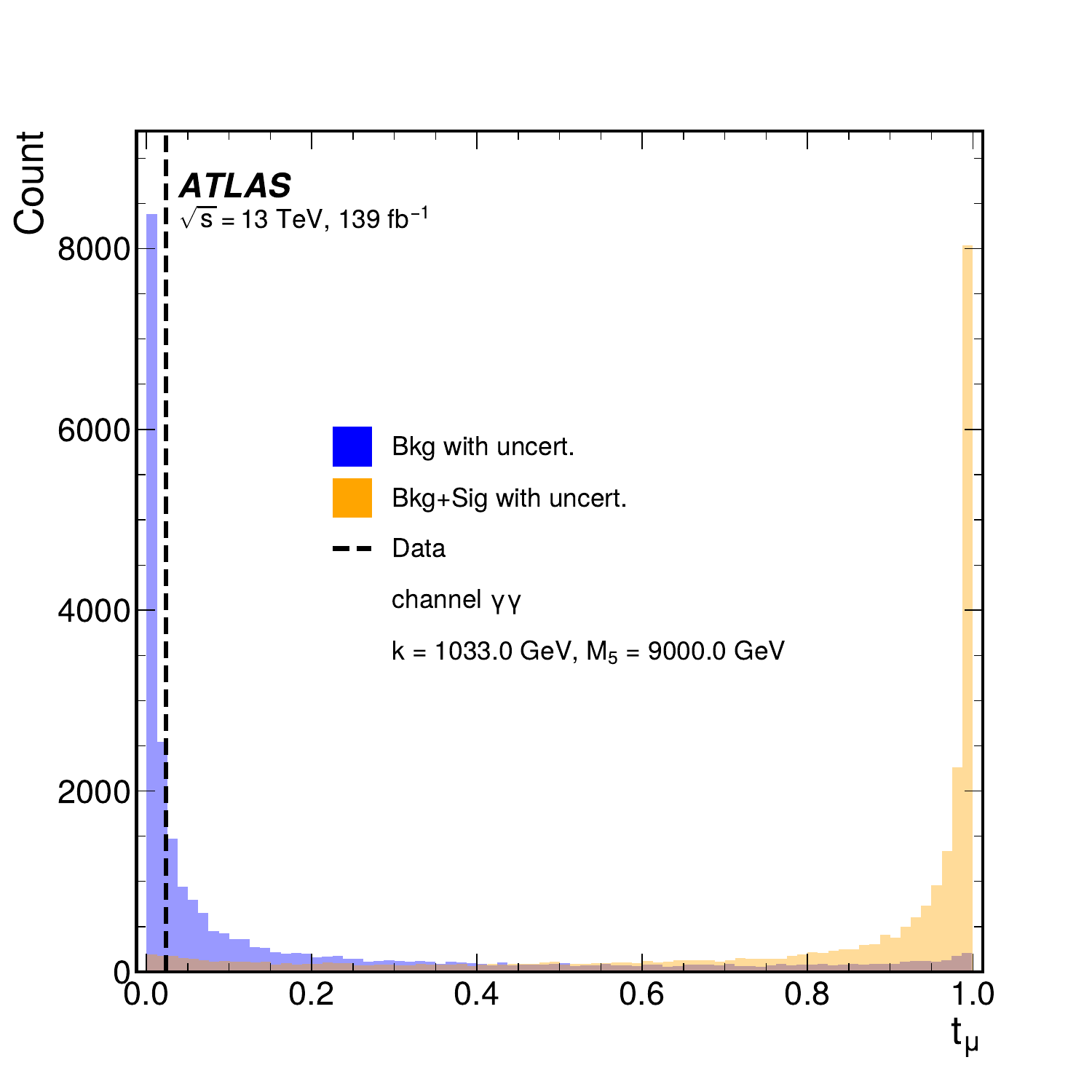}
\caption{The distribution of the test statistic
$t_{\mu}$ assuming a signal model $\mu$ with $k=1033$~\GeV and $M_{5}=9000$~\GeV and assuming the Standard Model in the diphoton channel for the case where mass thresholds are applied.
The pseudo-experiments used to generate the test statistic include the effects from the statistical uncertainty and all systematic uncertainties.
The test statistic from observed data is shown as a dotted line.
}
\label{fig:teststatdata}
\end{figure}
 
In the absence of a clear signal, limits at 95\% CL are set on the CW/LD model in the $k$--$M_{5}$ plane.
The limits for the case with mass thresholding, as discussed in Section~\ref{sSec:Classifier}, are shown in Figure~\ref{fig:CWLDExclusion}.
Additional exclusions are shown for the case without mass thresholds in Figure~\ref{fig:CWLDExclusionNoThreshold}.
As expected, the sensitivity and corresponding limits are stronger for the case without mass thresholds in the region of $k\lesssim1000$~\GeV.
 
\begin{figure}[htb!]
\centering
\subfloat[]{\includegraphics[width=0.49\textwidth]{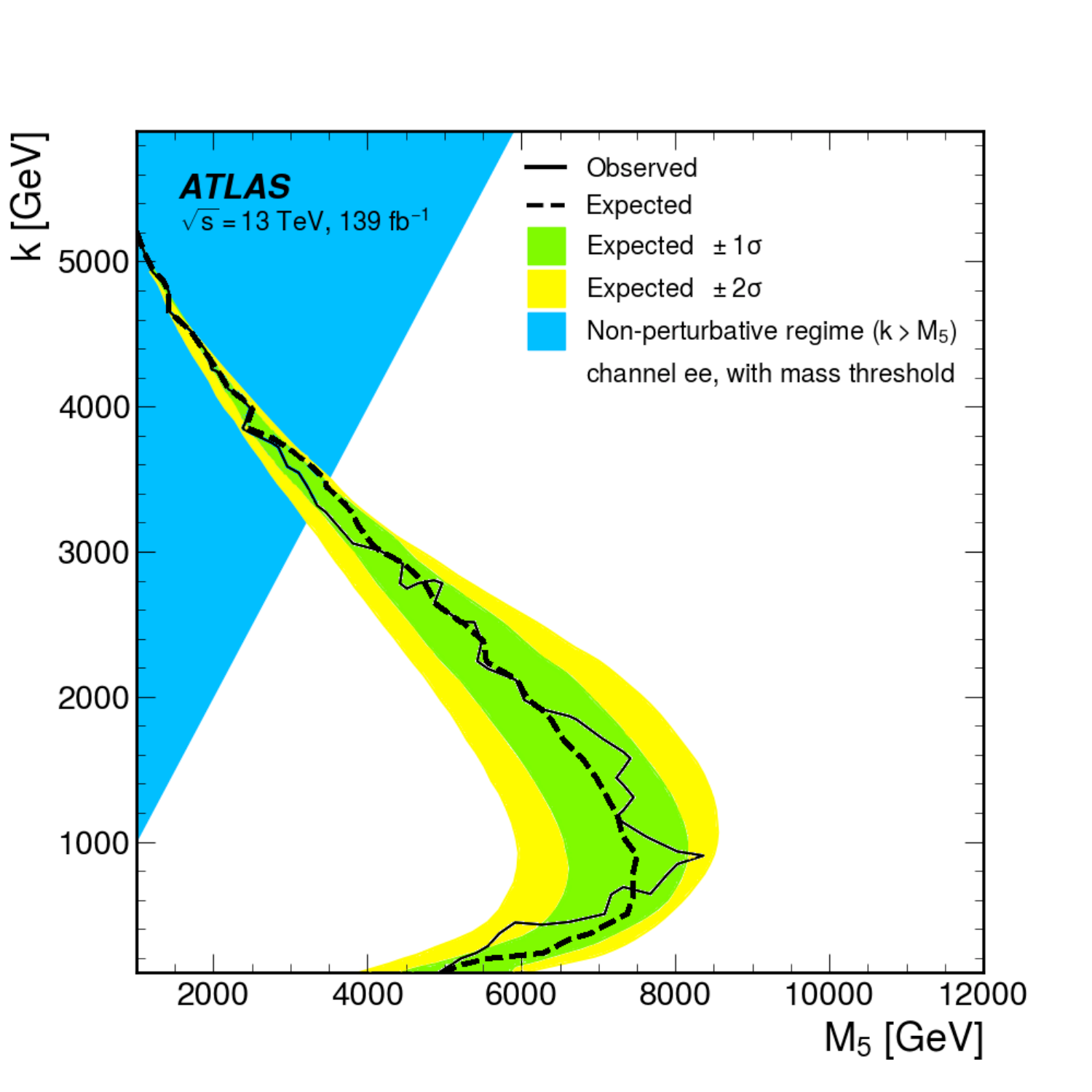}}
\subfloat[]{\includegraphics[width=0.49\textwidth]{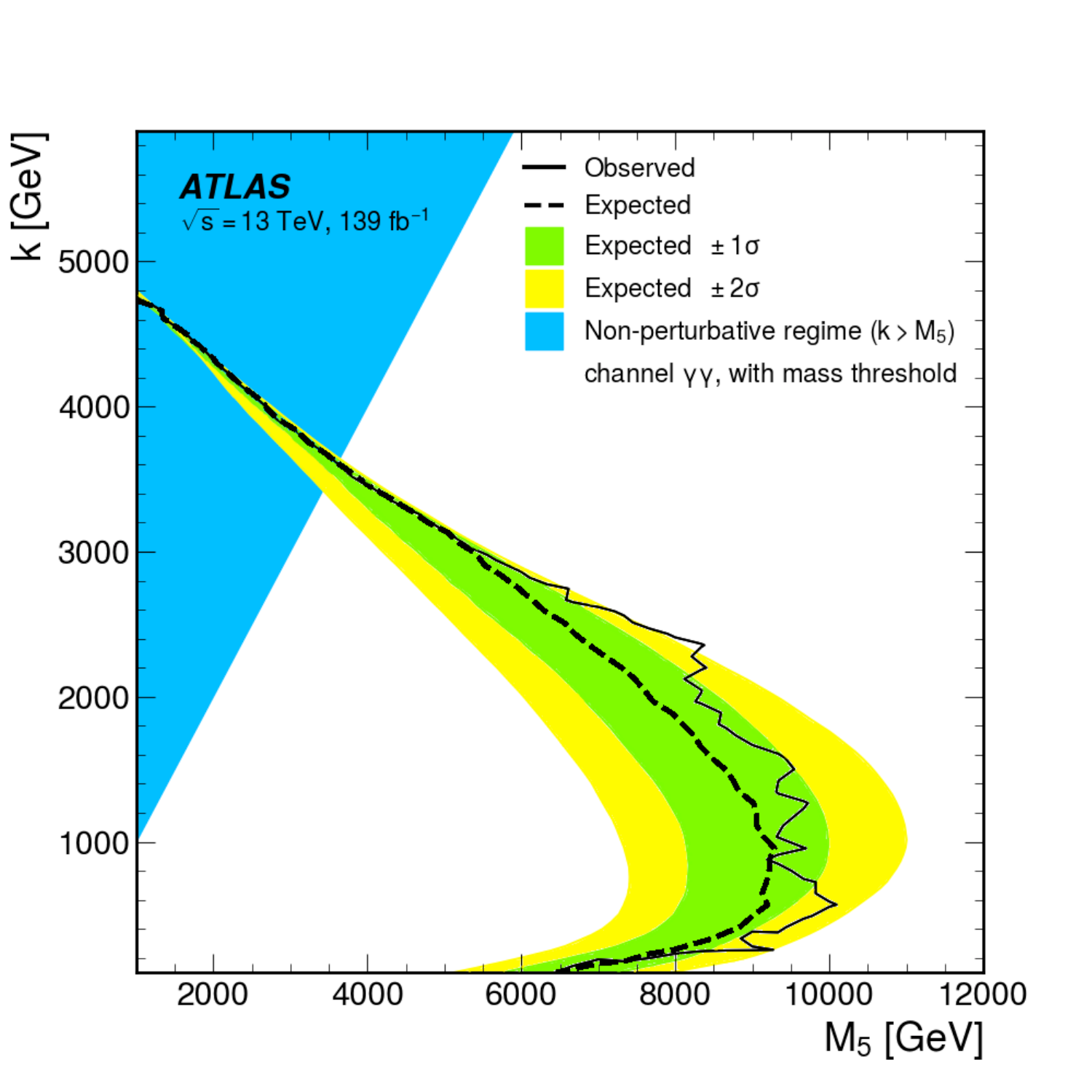}}
\caption{The expected and observed exclusion limits at 95\% CL for the clockwork gravity model projected in the $k$--$M_{5}$ parameter space for the (a) ee and (b) $\gamma\gamma$ channels, both for the case with mass thresholds.
The surrounding shaded bands represent the $\pm 1\sigma$ and $\pm 2\sigma$ uncertainties on the expected limit.
The shaded area with $k > M_{5}$ illustrates the region of parameter space where the CW/LD theory becomes non-perturbative.
}
\label{fig:CWLDExclusion}
\end{figure}
\begin{figure}[htb]
\centering
\subfloat[]{\includegraphics[width=0.49\textwidth]{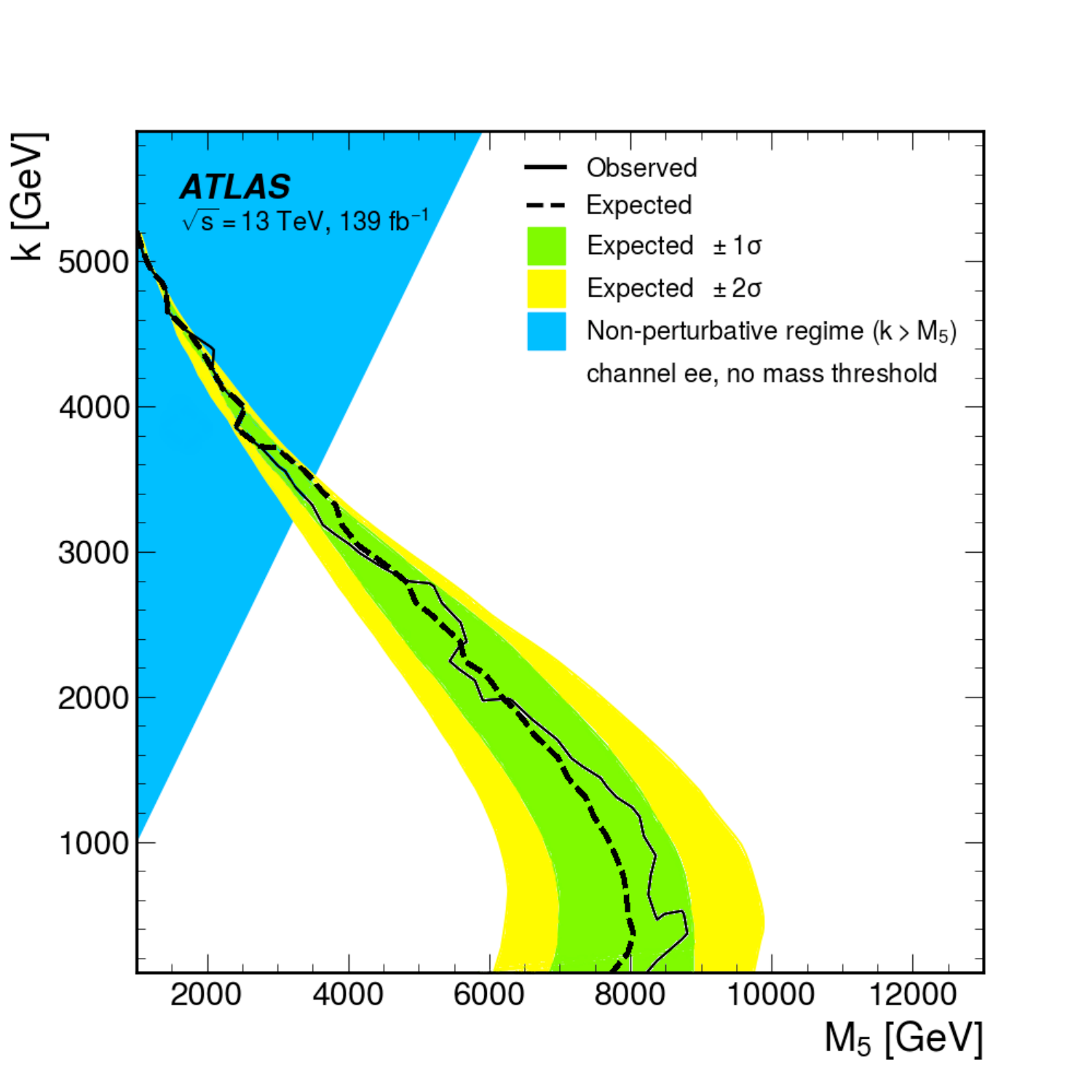}}
\subfloat[]{\includegraphics[width=0.49\textwidth]{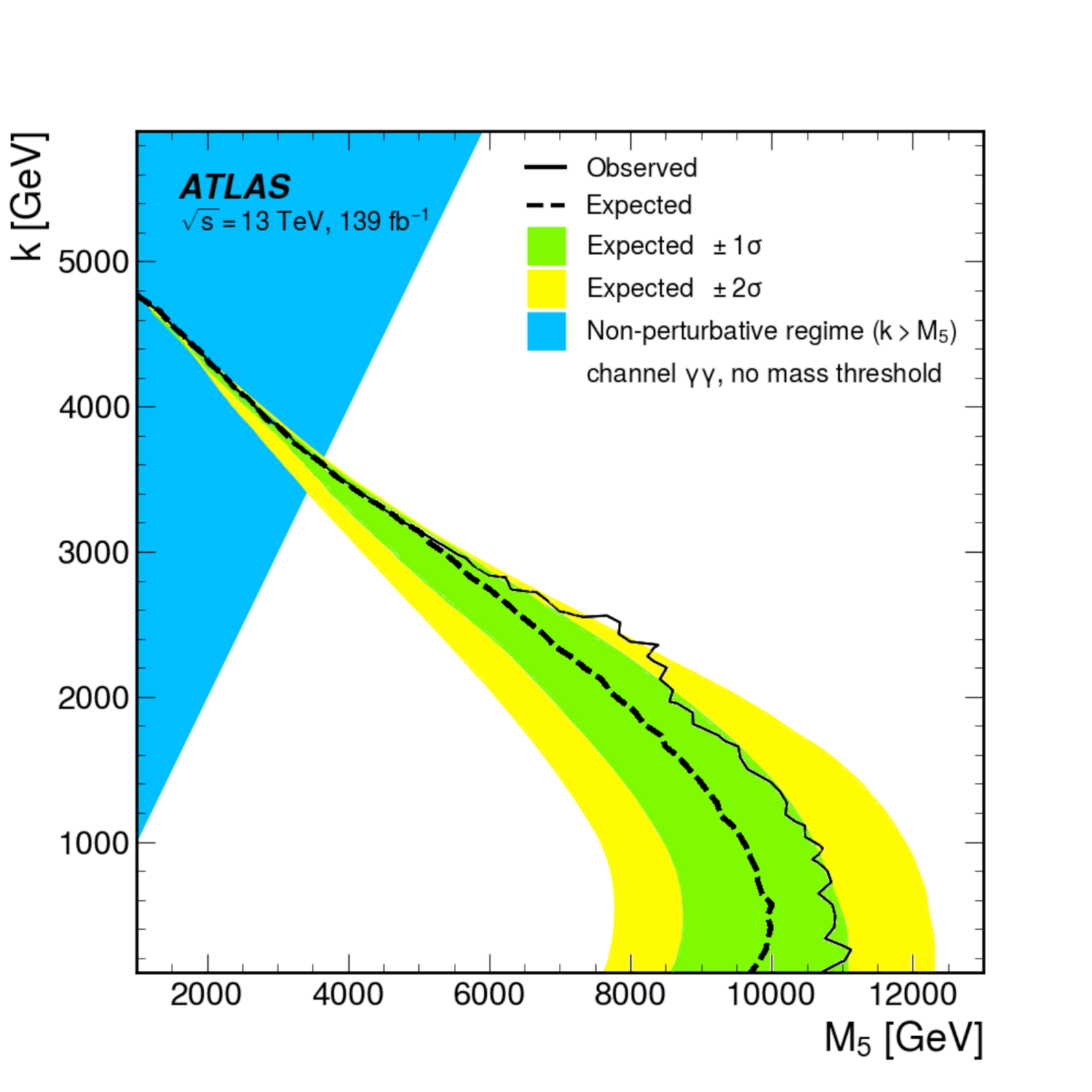}}
\caption{The expected and observed exclusion limits at 95\% CL for the clockwork gravity model projected in the $k$--$M_{5}$ parameter space for the (a) ee and (b) $\gamma\gamma$ channels, both for the case without mass thresholds.
The surrounding shaded bands represent the $\pm 1\sigma$ and $\pm 2\sigma$ uncertainties on the expected limit.
The shaded area with $k > M_{5}$ illustrates the region of phase space where the CW/LD theory becomes non-perturbative.
}
\label{fig:CWLDExclusionNoThreshold}
\end{figure}
 
For the case with mass thresholds, the maximum excluded $M_{5}$ value in the diphoton channel is approximately 10~TeV for values of $k\approx 600$~\GeV, while the maximum excluded $M_{5}$ value in the dielectron channel is approximately 8.4 TeV for values of $k\approx 900$~\GeV.
For the case without mass thresholds, the maximum excluded $M_{5}$ value in the diphoton channel is approximately 11 TeV for values of $k\approx 200-1000$~\GeV, while the maximum excluded $M_{5}$ value in the dielectron channel is approximately 8.8 TeV for values of $k\approx 400$~\GeV.
The areas in $k$--$M_{5}$ where the observed limits are stronger than the expected limits indicate that the data is effectively smoother than the expected fluctuations arising from the statistical and systematic uncertainties.
 
The systematic uncertainties mainly impact the classifier result in the range of $500<k<1500$~\GeV for both of the analysis channels. The statistical uncertainties dominate in the higher $k$ ranges.
The sensitivity of the classifier with systematic uncertainties is at most $\sim 1$~\TeV weaker in $M_5$ exclusion than the limits evaluated without including the systematic uncertainties.
The most dominant uncertainty contribution in both of the channels is due to the theoretical uncertainties in the background modelling.
 
It is worthwhile to mention that the NNs are initially trained on a lattice of points in the $k$--$M_{5}$ plane, where a dedicated training is performed for each lattice point.
The results are verified to be similarly effective for the models in-between those lattice points.


\FloatBarrier
 
\section{Conclusion}

Searches for periodic new phenomena in the dielectron and diphoton invariant mass spectra with the ATLAS experiment at the LHC are presented.
The searches are conducted with 139~\ifb\ of $pp$ collision data at $\sqrt{s}$ = 13~\TeV.
The model-dependent result is optimised for spin-2 periodic resonances (KK gravitons) in the clockwork/linear-dilaton model.
Model-independent results sensitive to more general periodic contributions are also presented.
 
The data observed in the $ee$ and $\gamma\gamma$ invariant mass spectra are found to be consistent with the SM expectations.
Without a significant excess, the $95\%$ CL exclusion limits are set in the parameter space of the CW/LD model.
This result excludes values of $M_{5}$ in the range 11~\TeV to 1~\TeV for values of $k$ in the range 100~\GeV to 5~\TeV.


\section*{Acknowledgements}


We thank CERN for the very successful operation of the LHC, as well as the
support staff from our institutions without whom ATLAS could not be
operated efficiently.
 
We acknowledge the support of
ANPCyT, Argentina;
YerPhI, Armenia;
ARC, Australia;
BMWFW and FWF, Austria;
ANAS, Azerbaijan;
CNPq and FAPESP, Brazil;
NSERC, NRC and CFI, Canada;
CERN;
ANID, Chile;
CAS, MOST and NSFC, China;
Minciencias, Colombia;
MEYS CR, Czech Republic;
DNRF and DNSRC, Denmark;
IN2P3-CNRS and CEA-DRF/IRFU, France;
SRNSFG, Georgia;
BMBF, HGF and MPG, Germany;
GSRI, Greece;
RGC and Hong Kong SAR, China;
ISF and Benoziyo Center, Israel;
INFN, Italy;
MEXT and JSPS, Japan;
CNRST, Morocco;
NWO, Netherlands;
RCN, Norway;
MEiN, Poland;
FCT, Portugal;
MNE/IFA, Romania;
MESTD, Serbia;
MSSR, Slovakia;
ARRS and MIZ\v{S}, Slovenia;
DSI/NRF, South Africa;
MICINN, Spain;
SRC and Wallenberg Foundation, Sweden;
SERI, SNSF and Cantons of Bern and Geneva, Switzerland;
MOST, Taiwan;
TENMAK, T\"urkiye;
STFC, United Kingdom;
DOE and NSF, United States of America.
In addition, individual groups and members have received support from
BCKDF, CANARIE, Compute Canada and CRC, Canada;
PRIMUS 21/SCI/017 and UNCE SCI/013, Czech Republic;
COST, ERC, ERDF, Horizon 2020 and Marie Sk{\l}odowska-Curie Actions, European Union;
Investissements d'Avenir Labex, Investissements d'Avenir Idex and ANR, France;
DFG and AvH Foundation, Germany;
Herakleitos, Thales and Aristeia programmes co-financed by EU-ESF and the Greek NSRF, Greece;
BSF-NSF and MINERVA, Israel;
Norwegian Financial Mechanism 2014-2021, Norway;
NCN and NAWA, Poland;
La Caixa Banking Foundation, CERCA Programme Generalitat de Catalunya and PROMETEO and GenT Programmes Generalitat Valenciana, Spain;
G\"{o}ran Gustafssons Stiftelse, Sweden;
The Royal Society and Leverhulme Trust, United Kingdom.
 
The crucial computing support from all WLCG partners is acknowledged gratefully, in particular from CERN, the ATLAS Tier-1 facilities at TRIUMF (Canada), NDGF (Denmark, Norway, Sweden), CC-IN2P3 (France), KIT/GridKA (Germany), INFN-CNAF (Italy), NL-T1 (Netherlands), PIC (Spain), ASGC (Taiwan), RAL (UK) and BNL (USA), the Tier-2 facilities worldwide and large non-WLCG resource providers. Major contributors of computing resources are listed in Ref.~\cite{ATL-SOFT-PUB-2021-003}.


\printbibliography

@article{Abadi:2016kic,
      author         = "Abadi, Martín and others",
      title          = "{TensorFlow: Large-Scale Machine Learning on
                        Heterogeneous Distributed  Systems}",
      year           = "2016",
      eprint         = "1603.04467",
      archivePrefix  = "arXiv",
      primaryClass   = "cs.DC",
      SLACcitation   = "%%CITATION = ARXIV:1603.04467;%%"
}

@article{Kingma:2014vow,
      author         = "Kingma, Diederik P. and Ba, Jimmy",
      title          = "{Adam: A Method for Stochastic Optimization}",
      year           = "2014",
      eprint         = "1412.6980",
      archivePrefix  = "arXiv",
      primaryClass   = "cs.LG",
      SLACcitation   = "%%CITATION = ARXIV:1412.6980;%%"
}

@article{Farina:2018fyg,
    author = "Farina, Marco and Nakai, Yuichiro and Shih, David",
    title = "{Searching for New Physics with Deep Autoencoders}",
    eprint = "1808.08992",
    archivePrefix = "arXiv",
    primaryClass = "hep-ph",
    doi = "10.1103/PhysRevD.101.075021",
    journal = "Phys. Rev. D",
    volume = "101",
    number = "7",
    pages = "075021",
    year = "2020"
}

@article{Giudice:2017fmj,
      author         = "Giudice, Gian F. and Kats, Yevgeny and McCullough,
                        Matthew and Torre, Riccardo and Urbano, Alfredo",
      title          = "{Clockwork/linear dilaton: structure and phenomenology}",
      journal        = "JHEP",
      volume         = "06",
      year           = "2018",
      pages          = "009",
      doi            = "10.1007/JHEP06(2018)009",
      eprint         = "1711.08437",
      archivePrefix  = "arXiv",
      primaryClass   = "hep-ph",
      reportNumber   = "CERN-TH-2017-219",
      SLACcitation   = "%%CITATION = ARXIV:1711.08437;%%"
}

@article{Randall:1999ee,
      author         = "Randall, Lisa and Sundrum, Raman",
      title          = "{Large mass hierarchy from a small extra dimension}",
      journal        = "Phys. Rev. Lett.",
      volume         = "83",
      year           = "1999",
      pages          = "3370-3373",
      doi            = "10.1103/PhysRevLett.83.3370",
      eprint         = "hep-ph/9905221",
      archivePrefix  = "arXiv",
      reportNumber   = "MIT-CTP-2860, PUPT-1860, BUHEP-99-9",
      SLACcitation   = "%%CITATION = HEP-PH/9905221;%%"
}

@article{Antoniadis:2001sw,
	author         = "Antoniadis, Ignatios and Dimopoulos, Savas and Giveon,
	Amit",
	title          = "{Little string theory at a TeV}",
	journal        = "JHEP",
	volume         = "05",
	year           = "2001",
	pages          = "055",
	doi            = "10.1088/1126-6708/2001/05/055",
	eprint         = "hep-th/0103033",
	archivePrefix  = "arXiv",
	reportNumber   = "CERN-TH-2001-066, RI-07-00, ITP-01-04",
	SLACcitation   = "%%CITATION = HEP-TH/0103033;%%"
}

@article{Berkooz:1997cq,
	author         = "Berkooz, Micha and Rozali, Moshe and Seiberg, Nathan",
	title          = "{Matrix description of M-theory on $T^4$ and $T^5$}",
	journal        = "Phys. Lett.",
	volume         = "B408",
	year           = "1997",
	pages          = "105-110",
	doi            = "10.1016/S0370-2693(97)00800-9",
	eprint         = "hep-th/9704089",
	archivePrefix  = "arXiv",
	reportNumber   = "UTTG-12-97, RU-97-23",
	SLACcitation   = "%%CITATION = HEP-TH/9704089;%%"
}

@article{Seiberg:1997zk,
	author         = "Seiberg, Nathan",
	title          = "{New theories in six-dimensions and matrix description of
	M-theory on $T^5$ and $T^5/Z_2$}",
	journal        = "Phys. Lett.",
	volume         = "B408",
	year           = "1997",
	pages          = "98-104",
	doi            = "10.1016/S0370-2693(97)00805-8",
	eprint         = "hep-th/9705221",
	archivePrefix  = "arXiv",
	reportNumber   = "RU-97-42",
	SLACcitation   = "%%CITATION = HEP-TH/9705221;%%"
}

@article{Giudice:2016yja,
	author         = "Giudice, Gian F. and McCullough, Matthew",
	title          = "{A Clockwork Theory}",
	journal        = "JHEP",
	volume         = "02",
	year           = "2017",
	pages          = "036",
	doi            = "10.1007/JHEP02(2017)036",
	eprint         = "1610.07962",
	archivePrefix  = "arXiv",
	primaryClass   = "hep-ph",
	reportNumber   = "CERN-TH-2016-223",
	SLACcitation   = "%%CITATION = ARXIV:1610.07962;%%"
}

@article{Antoniadis:2011qw,
	author         = "Antoniadis, Ignatios and Arvanitaki, Asimina and
	Dimopoulos, Savas and Giveon, Amit",
	title          = "{Phenomenology of TeV Little String Theory from
	Holography}",
	journal        = "Phys. Rev. Lett.",
	volume         = "108",
	year           = "2012",
	pages          = "081602",
	doi            = "10.1103/PhysRevLett.108.081602",
	eprint         = "1102.4043",
	archivePrefix  = "arXiv",
	primaryClass   = "hep-ph",
	reportNumber   = "CERN-PH-TH-2011-024",
	SLACcitation   = "%%CITATION = ARXIV:1102.4043;%%"
}

@article{Martin:2004dh,
    author = "Martin, A. D. and Roberts, R. G. and Stirling, W. J. and Thorne, R. S.",
    title = "{Parton distributions incorporating QED contributions}",
    eprint = "hep-ph/0411040",
    archivePrefix = "arXiv",
    reportNumber = "IPPP-04-62, DCPT-04-124, CAVENDISH-HEP-2004-28",
    doi = "10.1140/epjc/s2004-02088-7",
    journal = "Eur. Phys. J. C",
    volume = "39",
    pages = "155--161",
    year = "2005"
}

@article{Cohen:2017exh,
	author         = "Cohen, Timothy and Freytsis, Marat and Ostdiek, Bryan",
	title          = "{(Machine) Learning to Do More with Less}",
	journal        = "JHEP",
	volume         = "02",
	year           = "2018",
	pages          = "034",
	doi            = "10.1007/JHEP02(2018)034",
	eprint         = "1706.09451",
	archivePrefix  = "arXiv",
	primaryClass   = "hep-ph",
	SLACcitation   = "%%CITATION = ARXIV:1706.09451;%%"
}

@article{Heimel:2018mkt,
      author         = "Heimel, Theo and Kasieczka, Gregor and Plehn, Tilman and
                        Thompson, Jennifer M.",
      title          = "{QCD or What?}",
      journal        = "SciPost Phys.",
      volume         = "6",
      year           = "2019",
      number         = "3",
      pages          = "030",
      doi            = "10.21468/SciPostPhys.6.3.030",
      eprint         = "1808.08979",
      archivePrefix  = "arXiv",
      primaryClass   = "hep-ph",
      SLACcitation   = "%%CITATION = ARXIV:1808.08979;%%"
}

@article{ArkaniHamed:1998rs,
    author = "Arkani-Hamed, Nima and Dimopoulos, Savas and Dvali, G.R.",
    title = "{The Hierarchy problem and new dimensions at a millimeter}",
    eprint = "hep-ph/9803315",
    archivePrefix = "arXiv",
    reportNumber = "SLAC-PUB-7769, SU-ITP-98-13",
    doi = "10.1016/S0370-2693(98)00466-3",
    journal = "Phys. Lett. B",
    volume = "429",
    pages = "263--272",
    year = "1998"
}

@article{Antoniadis:1998ig,
    author = "Antoniadis, Ignatios and Arkani-Hamed, Nima and Dimopoulos, Savas and Dvali, G. R.",
    title = "{New dimensions at a millimeter to a Fermi and superstrings at a TeV}",
    eprint = "hep-ph/9804398",
    archivePrefix = "arXiv",
    reportNumber = "SLAC-PUB-7801, SU-ITP-98-28, CPTH-S608-0498, IC-98-39",
    doi = "10.1016/S0370-2693(98)00860-0",
    journal = "Phys. Lett. B",
    volume = "436",
    pages = "257--263",
    year = "1998"
}

@article{Beauchesne:2019tpx,
    author = "Beauchesne, Hugues and Kats, Yevgeny",
    title = "{Searching for periodic signals in kinematic distributions using continuous wavelet transforms}",
    eprint = "1907.03676",
    archivePrefix = "arXiv",
    primaryClass = "hep-ph",
    doi = "10.1140/epjc/s10052-020-7746-8",
    journal = "Eur. Phys. J. C",
    volume = "80",
    number = "3",
    pages = "192",
    year = "2020"
}

@phdthesis{Oreglia:1980cs,
    author = "Oreglia, M.",
    title = "A Study of the Reactions $\psi^\prime \to \gamma \gamma \psi$",
    school = "Stanford University",
    year = "1980",
    note = "SLAC-236"
}

@phdthesis{Skwarnicki:1986xj,
    author = "Skwarnicki, Tomasz",
    title = "A study of the radiative CASCADE transitions between the Upsilon-Prime and Upsilon resonances",
    school = "Cracow Institut of Nuclear Physics",
    year = "1986",
    note = "DESY-F31-86-02"
}

@article{Oreglia:1981fx,
  title = {Study of the reaction ${\ensuremath{\psi}}^{\ensuremath{'}}\ensuremath{\rightarrow}\ensuremath{\gamma}\ensuremath{\gamma}\frac{J}{\ensuremath{\psi}}$},
  author = {Oreglia, M. and Bloom, E. and Bulos, F. and Chestnut, R. and Gaiser, J. and Godfrey, G. and Kiesling, C. and Lockman, W. and Scharre, D. L. and Partridge, R. and Peck, C. and Porter, F. C. and Antreasyan, D. and Gu, Yi-Fan and Kollmann, W. and Richardson, M. and Strauch, K. and Wacker, K. and Aschman, D. and Burnett, T. and Newman, C. and Cavalli-Sforza, M. and Coyne, D. and Sadrozinski, H. and Hofstadter, R. and Kirkbride, I. and Kolanoski, H. and K\"onigsmann, K. and Liberman, A. and O'Reilly, J. and Pollock, B. and Tompkins, J.},
  journal = {Phys. Rev. D},
  volume = {25},
  issue = {9},
  pages = {2259--2277},
  numpages = {0},
  year = {1982},
  month = {5},
  publisher = {American Physical Society},
  doi = {10.1103/PhysRevD.25.2259},
  url = {https://link.aps.org/doi/10.1103/PhysRevD.25.2259}
}

@article{Avoni:2018iuv,
author = "Avoni, G. and others",
    title = "{The new LUCID-2 detector for luminosity measurement and monitoring in ATLAS}",
    doi = "10.1088/1748-0221/13/07/P07017",
    journal = "JINST",
    volume = "13",
    number = "07",
    pages = "P07017",
    year = "2018"
}

@article{Junk:1999kv,
    author = "Junk, Thomas",
    title = "{Confidence level computation for combining searches with small statistics}",
    eprint = "hep-ex/9902006",
    archivePrefix = "arXiv",
    reportNumber = "CARLETON-OPAL-PHYS-99-01, CERN-EP-99-041",
    doi = "10.1016/S0168-9002(99)00498-2",
    journal = "Nucl. Instrum. Meth. A",
    volume = "434",
    pages = "435--443",
    year = "1999"
}

@article{ParticleDataGroup:2020ssz,
    author = "{Particle Data Group} and Zyla, P.A. and others",
    %collaboration = "Particle Data Group",
    title = "{Review of Particle Physics}",
    doi = "10.1093/ptep/ptaa104",
    journal = "PTEP",
    volume = "2020",
    number = "8",
    pages = "083C01",
    year = "2020"
}

@article{Arbuzov:2015yja,
    author = "Arbuzov, A. and Bardin, D. and Bondarenko, S. and Christova, P. and Kalinovskaya, L. and Klein, U. and Kolesnikov, V. and Rumyantsev, L. and Sadykov, R. and Sapronov, A.",
    title = "{Update of the MCSANC Monte Carlo integrator, v. 1.20}",
    eprint = "1509.03052",
    archivePrefix = "arXiv",
    primaryClass = "hep-ph",
    doi = "10.1134/S0021364016020041",
    journal = "JETP Lett.",
    volume = "103",
    number = "2",
    pages = "131--136",
    year = "2016"
}

@article{ParticleDataGroup:2018ovx,
    author = "{Particle Data Group} and Tanabashi, M. and others",
    %collaboration = "Particle Data Group",
    title = "{Review of Particle Physics}",
    doi = "10.1103/PhysRevD.98.030001",
    journal = "Phys. Rev. D",
    volume = "98",
    number = "3",
    pages = "030001",
    year = "2018"
}

@article{Morlet,
        title = {Wave propagation and sampling theory—Part I: Complex signal and scattering in multilayered media },
        author = {Morlet, J. and Arens,G. and Fourgeau,E. and Giard,D. },
        journal={ Geophysics },
        year={ 1982 },
        publisher={ Society of Exploration Geophysicists },
        volume={ 47 },
        pages={ 203-221 },
        number={ 2 },
        doi={ 10.1190/1.1441328 },  
      }

@article{COHEN201981,
      Author = {Michael X Cohen},
      Doi = {https://doi.org/10.1016/j.neuroimage.2019.05.048},
      Issn = {1053-8119},
      Journal = {NeuroImage},
      Pages = {81-86},
      Title = {A better way to define and describe Morlet wavelets for time-frequency analysis},
      Url = {https://www.sciencedirect.com/science/article/pii/S1053811919304409},
      Volume = {199},
      Year = {2019},
      }

@article{Lai:2010vv,
      author         = "Lai, H.-L. and others",
      title          = "{New parton distributions for collider physics}",
      journal        = "Phys. Rev. D",
      volume         = "82",
      pages          = "074024",
      doi            = "10.1103/PhysRevD.82.074024",
      year           = "2010",
      eprint         = "1007.2241",
      archivePrefix  = "arXiv",
      primaryClass   = "hep-ph",
      reportNumber   = "MSUHEP-100707, SMU-HEP-10-10",
      SLACcitation   = "%%CITATION = ARXIV:1007.2241;%%",
}

@article{Dulat:2015mca,
      author         = "Dulat, Sayipjamal and Hou, Tie-Jiun and Gao, Jun and
                        Guzzi, Marco and Huston, Joey and Nadolsky, Pavel and
                        Pumplin, Jon and Schmidt, Carl and Stump, Daniel and Yuan,
                        C. P.",
      title          = "{New parton distribution functions from a global analysis
                        of quantum chromodynamics}",
      journal        = "Phys. Rev. D",
      volume         = "93",
      year           = "2016",
      number         = "3",
      pages          = "033006",
      doi            = "10.1103/PhysRevD.93.033006",
      eprint         = "1506.07443",
      archivePrefix  = "arXiv",
      primaryClass   = "hep-ph",
      SLACcitation   = "%%CITATION = ARXIV:1506.07443;%%"
}

@article{Ball:2012cx,
      author         = "{NNPDF Collaboration} and Ball, Richard D. and others",
      title          = "{Parton distributions with LHC data}",
      journal        = "Nucl. Phys. B",
      volume         = "867",
      year           = "2013",
      pages          = "244",
      doi            = "10.1016/j.nuclphysb.2012.10.003",
      eprint         = "1207.1303",
      archivePrefix  = "arXiv",
      primaryClass   = "hep-ph",
      reportNumber   = "EDINBURGH-2012-08, IFUM-FT-997, FR-PHENO-2012-014,
                        RWTH-TTK-12-25, CERN-PH-TH-2012-037, SFB-CPP-12-47\,
                        --CERN-PH-TH-2012-037",
      SLACcitation   = "%%CITATION = ARXIV:1207.1303;%%"
}

@article{Ball:2014uwa,
      author         = "{The NNPDF Collaboration} and Ball, Richard D. and others",
      title          = "{Parton distributions for the LHC run II}",
      journal        = "JHEP",
      volume         = "04",
      year           = "2015",
      pages          = "040",
      doi            = "10.1007/JHEP04(2015)040",
      eprint         = "1410.8849",
      archivePrefix  = "arXiv",
      primaryClass   = "hep-ph",
      reportNumber   = "EDINBURGH-2014-15, IFUM-1034-FT, CERN-PH-TH-2013-253,
                        OUTP-14-11P, CAVENDISH-HEP-14-11",
      SLACcitation   = "%%CITATION = ARXIV:1410.8849;%%"
}

@Article{Sjostrand:2007gs,
	author    = "Sj{\"o}strand, T. and Mrenna, S. and Skands, P.",
	title     = "{A brief introduction to PYTHIA 8.1}",
	journal   = "Comput. Phys. Commun.",
	volume    = "178",
	year      = "2008",
	pages     = "852-867",
	eprint    = "0710.3820",
	archivePrefix = "arXiv",
	primaryClass  =  "hep-ph",
	doi       = "10.1016/j.cpc.2008.01.036",
	SLACcitation  = "%%CITATION = 0710.3820;%%"
}

@article{Sjostrand:2014zea,
      author         = "Sj{\"o}strand, Torbj{\"o}rn and Ask, Stefan and Christiansen,
                        Jesper R. and Corke, Richard and Desai, Nishita and Ilten,
                        Philip and Mrenna, Stephen and Prestel, Stefan and
                        Rasmussen, Christine O. and Skands, Peter Z.",
      title          = "{An introduction to PYTHIA 8.2}",
      journal        = "Comput. Phys. Commun.",
      volume         = "191",
      year           = "2015",
      pages          = "159",
      doi            = "10.1016/j.cpc.2015.01.024",
      eprint         = "1410.3012",
      archivePrefix  = "arXiv",
      primaryClass   = "hep-ph",
      reportNumber   = "LU-TP-14-36, MCNET-14-22, CERN-PH-TH-2014-190,
                        FERMILAB-PUB-14-316-CD, DESY-14-178, SLAC-PUB-16122,
                        --FERMILAB-PUB-14-316-CD",
      SLACcitation   = "%%CITATION = ARXIV:1410.3012;%%"
}

@Article{Nason:2004rx,
      author         = "Nason, Paolo",
      title          = "{A new method for combining NLO QCD with shower Monte Carlo algorithms}",
      journal        = "JHEP",
      volume         = "11",
      pages          = "040",
      doi            = "10.1088/1126-6708/2004/11/040",
      year           = "2004",
      eprint         = "hep-ph/0409146",
      archivePrefix  = "arXiv",
}

@Article{Frixione:2007vw,
      author         = "Frixione, Stefano and Nason, Paolo and Oleari, Carlo",
      title          = "{Matching NLO QCD computations with parton shower
                        simulations: the POWHEG method}",
      journal        = "JHEP",
      volume         = "11",
      pages          = "070",
      doi            = "10.1088/1126-6708/2007/11/070",
      year           = "2007",
      eprint         = "0709.2092",
      archivePrefix  = "arXiv",
      primaryClass   = "hep-ph",
}

@Article{Alioli:2010xd,
      author         = "Alioli, Simone and Nason, Paolo and Oleari, Carlo and Re,
                        Emanuele",
      title          = "{A general framework for implementing NLO calculations in
                        shower Monte Carlo programs: the POWHEG BOX}",
      journal        = "JHEP",
      volume         = "06",
      pages          = "043",
      doi            = "10.1007/JHEP06(2010)043",
      year           = "2010",
      eprint         = "1002.2581",
      archivePrefix  = "arXiv",
      primaryClass   = "hep-ph",
}

@article{Alioli:2008gx,
      author         = "Alioli, Simone and Nason, Paolo and Oleari, Carlo and Re,
                        Emanuele",
      title          = "{NLO vector-boson production matched with shower in
                        POWHEG}",
      journal        = "JHEP",
      volume         = "07",
      year           = "2008",
      pages          = "060",
      doi            = "10.1088/1126-6708/2008/07/060",
      eprint         = "0805.4802",
      archivePrefix  = "arXiv",
      primaryClass   = "hep-ph",
      reportNumber   = "NSF-KITP-08-74",
      SLACcitation   = "%%CITATION = ARXIV:0805.4802;%%"
}

@article{Alioli:2009je,
      author         = "Alioli, Simone and Nason, Paolo and Oleari, Carlo and Re,
                        Emanuele",
      title          = "{NLO single-top production matched with shower in POWHEG:
                        \(s\)- and \(t\)-channel contributions}",
      journal        = "JHEP",
      volume         = "09",
      year           = "2009",
      pages          = "111",
      doi            = "10.1088/1126-6708/2009/09/111",
      eprint         = "0907.4076",
      archivePrefix  = "arXiv",
      primaryClass   = "hep-ph",
      SLACcitation   = "%%CITATION = ARXIV:0907.4076;%%",
      related        = "Alioli:2009je-err",
      relatedstring  = "Erratum:",
}

@article{Frederix:2012dh,
      author         = "Frederix, Rikkert and Re, Emanuele and Torrielli, Paolo",
      title          = "{Single-top \(t\)-channel hadroproduction in the four-flavour
                        scheme with POWHEG and aMC@NLO}",
      journal        = "JHEP",
      volume         = "09",
      year           = "2012",
      pages          = "130",
      doi            = "10.1007/JHEP09(2012)130",
      eprint         = "1207.5391",
      archivePrefix  = "arXiv",
      primaryClass   = "hep-ph",
      reportNumber   = "ZU-TH-14-12, IPPP-12-54, DCPT-12-108,
                        CERN-PH-TH-2012-206, MCNET-12-10, LPN12-084",
      SLACcitation   = "%%CITATION = ARXIV:1207.5391;%%"
}

@article{Re:2010bp,
      author         = "Re, Emanuele",
      title          = "{Single-top \(Wt\)-channel production matched with parton
                        showers using the POWHEG method}",
      journal        = "Eur. Phys. J. C",
      volume         = "71",
      year           = "2011",
      pages          = "1547",
      doi            = "10.1140/epjc/s10052-011-1547-z",
      eprint         = "1009.2450",
      archivePrefix  = "arXiv",
      primaryClass   = "hep-ph",
      reportNumber   = "IPPP-10-74, DCPT-10-148",
      SLACcitation   = "%%CITATION = ARXIV:1009.2450;%%"
}

@article{Anastasiou:2003ds,
      author         = "Anastasiou, Charalampos and Dixon, Lance and Melnikov,
                        Kirill and Petriello, Frank",
      title          = "{High-precision QCD at hadron colliders: Electroweak
                        gauge boson rapidity distributions at next-to-next-to leading order}",
      journal        = "Phys. Rev. D",
      volume         = "69",
      year           = "2004",
      pages          = "094008",
      doi            = "10.1103/PhysRevD.69.094008",
      eprint         = "hep-ph/0312266",
      archivePrefix  = "arXiv",
      reportNumber   = "SLAC-PUB-10288, UH-511-1042-03",
      SLACcitation   = "%%CITATION = HEP-PH/0312266;%%"
}

@article{Frixione:2007nw,
      author         = "Frixione, Stefano and Ridolfi, Giovanni and Nason, Paolo",
      title          = "{A positive-weight next-to-leading-order Monte Carlo for
                        heavy flavour hadroproduction}",
      journal        = "JHEP",
      volume         = "09",
      pages          = "126",
      doi            = "10.1088/1126-6708/2007/09/126",
      year           = "2007",
      eprint         = "0707.3088",
      archivePrefix  = "arXiv",
      primaryClass   = "hep-ph",
      SLACcitation   = "%%CITATION = ARXIV:0707.3088;%%",
}

@Article{Gleisberg:2008ta,
      author         = "Gleisberg, T. and H{\"o}che, S. and Krauss, F. and Sch{\"o}nherr, M. and Schumann, S. and Siegert, F. and Winter J.",
      title          = "{Event generation with SHERPA 1.1}",
      journal        = "JHEP",
      volume         = "02",
      pages          = "007",
      doi            = "10.1088/1126-6708/2009/02/007",
      year           = "2009",
      eprint         = "0811.4622",
      archivePrefix  = "arXiv",
      primaryClass   = "hep-ph",
}

@Article{Hoeche:2009rj,
      author         = "H{\"o}che, Stefan and Krauss, Frank and Schumann, Steffen and Siegert, Frank",
      title          = "{QCD matrix elements and truncated showers}",
      journal        = "JHEP",
      volume         = "05",
      pages          = "053",
      doi            = "10.1088/1126-6708/2009/05/053",
      year           = "2009",
      eprint         = "0903.1219",
      archivePrefix  = "arXiv",
      primaryClass   = "hep-ph",
}

@Article{Gleisberg:2008fv,
      author         = "Gleisberg, Tanju and H{\"o}che, Stefan",
      title          = "{Comix, a new matrix element generator}",
      journal        = "JHEP",
      volume         = "12",
      pages          = "039",
      doi            = "10.1088/1126-6708/2008/12/039",
      year           = "2008",
      eprint         = "0808.3674",
      archivePrefix  = "arXiv",
      primaryClass   = "hep-ph",
}

@Article{Schumann:2007mg,
      author         = "Schumann, Steffen and Krauss, Frank",
      title          = "{A parton shower algorithm based on Catani--Seymour dipole factorisation}",
      journal        = "JHEP",
      volume         = "03",
      pages          = "038",
      doi            = "10.1088/1126-6708/2008/03/038",
      year           = "2008",
      eprint         = "0709.1027",
      archivePrefix  = "arXiv",
      primaryClass   = "hep-ph",
}

@article{Hoeche:2012yf,
      author         = "H{\"o}che, Stefan and Krauss, Frank and Sch{\"o}nherr, Marek and
                        Siegert, Frank",
      title          = "{QCD matrix elements + parton showers. The NLO case}",
      journal        = "JHEP",
      volume         = "04",
      year           = "2013",
      pages          = "027",
      doi            = "10.1007/JHEP04(2013)027",
      eprint         = "1207.5030",
      archivePrefix  = "arXiv",
      primaryClass   = "hep-ph",
      reportNumber   = "SLAC-PUB-15191, IPPP-12-52, DCPT-12-104, LPN12-081,
                        FR-PHENO-2012-017, MCNET-12-09, --FR-PHENO-2012-017",
      SLACcitation   = "%%CITATION = ARXIV:1207.5030;%%"
}

@article{Bothmann:2019yzt,
      author = "Bothmann, Enrico and others",
      title = "{Event generation with Sherpa 2.2}",
      journal = "SciPost Phys.",
      volume = "7",
      year = "2019",
      number = "3",
      pages = "034",
      doi = "10.21468/SciPostPhys.7.3.034",
      reportNumber = "FERMILAB-PUB-19-218-T, SLAC-PUB-17433, IPPP/19/42, MCNET-19-11",
      eprint = "1905.09127",
      archivePrefix = "arXiv",
      primaryClass = "hep-ph",
}

@article{Hoeche:2011fd,
      author         = "H{\"o}che, Stefan and Krauss, Frank and Sch{\"o}nherr, Marek and
                        Siegert, Frank",
      title          = "{A critical appraisal of NLO+PS matching methods}",
      journal        = "JHEP",
      volume         = "09",
      year           = "2012",
      pages          = "049",
      doi            = "10.1007/JHEP09(2012)049",
      eprint         = "1111.1220",
      archivePrefix  = "arXiv",
      primaryClass   = "hep-ph",
      reportNumber   = "SLAC-PUB-14661, IPPP-11-67, DCPT-11-134, LPN11-58,
                        FR-PHENO-2011-019, MCNET-11-24",
      SLACcitation   = "%%CITATION = ARXIV:1111.1220;%%"
}

@article{Catani:2001cc,
      author         = "Catani, S. and Krauss, F. and Webber, B. R. and Kuhn, R.",
      title          = "{QCD Matrix Elements + Parton Showers}",
      journal        = "JHEP",
      volume         = "11",
      year           = "2001",
      pages          = "063",
      doi            = "10.1088/1126-6708/2001/11/063",
      eprint         = "hep-ph/0109231",
      archivePrefix  = "arXiv",
      reportNumber   = "CERN-TH-2000-367, CAVENDISH-HEP-00-03",
      SLACcitation   = "%%CITATION = HEP-PH/0109231;%%"
}

@article{Czakon:2011xx,
  author         = "Czakon, Michal and Mitov, Alexander",
  title          = "{Top++: A program for the calculation of the top-pair cross-section at hadron colliders}",
  journal        = "Comput. Phys. Commun.",
  volume         = "185",
  year           = "2014",
  pages          = "2930",
  doi            = "10.1016/j.cpc.2014.06.021",
  eprint         = "1112.5675",
  archivePrefix  = "arXiv",
  primaryClass   = "hep-ph",
  reportNumber   = "CERN-PH-TH-2011-303, TTK-11-58",
}

@article{Agostinelli:2002hh,
      author         = "Agostinelli, S. and others",
      title          = "{\textsc{Geant4} -- a simulation toolkit}",
      journal        = "Nucl. Instrum. Meth. A",
      volume         = "506",
      year           = "2003",
      pages          = "250",
      doi            = "10.1016/S0168-9002(03)01368-8",
      reportNumber   = "SLAC-PUB-9350, FERMILAB-PUB-03-339",
      SLACcitation   = "%%CITATION = NUIMA,A506,250;%%"
}

@Article{Read:2002hq,
    Author = {Read, Alexander L.},
    Journal = {J. Phys. G},
    Pages = {2693},
    doi = "10.1088/0954-3899/28/10/313",
    Title = "{Presentation of search results: the \(CL_S\) technique}",
    Volume = {28},
    Year = {2002}
}

@online{chollet2015keras,
      author         = {Chollet, Fran\c{c}ois and others},
      title          = {Keras},
      year           = {2015},
      url            = {https://keras.io},
}

@Article{PERF-2007-01,
    author         = "{ATLAS Collaboration}",
    title          = "{The ATLAS Experiment at the CERN Large Hadron Collider}",
    journal        = "JINST",
    volume         = "3",
    year           = "2008",
    pages          = "S08003",
    doi            = "10.1088/1748-0221/3/08/S08003",
    primaryClass   = "hep-ex",
}

@Article{SOFT-2010-01,
    author         = "{ATLAS Collaboration}",
    title          = "{The ATLAS Simulation Infrastructure}",
    journal        = "Eur. Phys. J. C",
    volume         = "70",
    year           = "2010",
    pages          = "823",
    doi            = "10.1140/epjc/s10052-010-1429-9",
    eprint         = "1005.4568",
    archivePrefix  = "arXiv",
    primaryClass   = "physics.ins-det",
}

@Article{STDM-2012-05,
    author         = "{ATLAS Collaboration}",
    title          = "{Measurement of isolated-photon pair production in \(pp\) collisions at \(\sqrt{s} = 7\,\text{TeV}\) with the ATLAS detector}",
    journal        = "JHEP",
    volume         = "01",
    year           = "2013",
    pages          = "086",
    doi            = "10.1007/JHEP01(2013)086",
    reportNumber   = "CERN-PH-EP-2012-300",
    eprint         = "1211.1913",
    archivePrefix  = "arXiv",
    primaryClass   = "hep-ex",
}

@Article{HIGG-2013-08,
    author         = "{ATLAS Collaboration}",
    title          = "{Measurement of Higgs boson production in the diphoton decay channel in \(pp\) collisions at center-of-mass energies of \(7\) and \(8\,\text{TeV}\) with the ATLAS detector}",
    journal        = "Phys. Rev. D",
    volume         = "90",
    year           = "2014",
    pages          = "112015",
    doi            = "10.1103/PhysRevD.90.112015",
    reportNumber   = "CERN-PH-EP-2014-198",
    eprint         = "1408.7084",
    archivePrefix  = "arXiv",
    primaryClass   = "hep-ex",
}

@Article{STDM-2015-05,
    author         = "{ATLAS Collaboration}",
    title          = "{Measurement of the Inelastic Proton--Proton Cross Section at \(\sqrt{s} = 13\,\text{TeV}\) with the ATLAS Detector at the LHC}",
    journal        = "Phys. Rev. Lett.",
    volume         = "117",
    year           = "2016",
    pages          = "182002",
    doi            = "10.1103/PhysRevLett.117.182002",
    reportNumber   = "CERN-EP-2016-140",
    eprint         = "1606.02625",
    archivePrefix  = "arXiv",
    primaryClass   = "hep-ex",
}

@Article{EXOT-2016-05,
    author         = "{ATLAS Collaboration}",
    title          = "{Search for new high-mass phenomena in the dilepton final state using \(36\,\text{fb}^{-1}\) of proton--proton collision data at \(\sqrt{s} = 13\,\text{TeV}\) with the ATLAS detector}",
    journal        = "JHEP",
    volume         = "10",
    year           = "2017",
    pages          = "182",
    doi            = "10.1007/JHEP10(2017)182",
    reportNumber   = "CERN-EP-2017-119",
    eprint         = "1707.02424",
    archivePrefix  = "arXiv",
    primaryClass   = "hep-ex",
}

@Article{HIGG-2016-08,
    author         = "{ATLAS Collaboration}",
    title          = "{Search for resonances in diphoton events at \(\sqrt{s} = 13\,\text{TeV}\) with the ATLAS detector}",
    journal        = "JHEP",
    volume         = "09",
    year           = "2016",
    pages          = "001",
    doi            = "10.1007/JHEP09(2016)001",
    reportNumber   = "CERN-EP-2016-120",
    eprint         = "1606.03833",
    archivePrefix  = "arXiv",
    primaryClass   = "hep-ex",
}

@Article{TRIG-2016-01,
    author         = "{ATLAS Collaboration}",
    title          = "{Performance of the ATLAS trigger system in 2015}",
    journal        = "Eur. Phys. J. C",
    volume         = "77",
    year           = "2017",
    pages          = "317",
    doi            = "10.1140/epjc/s10052-017-4852-3",
    reportNumber   = "CERN-EP-2016-241",
    eprint         = "1611.09661",
    archivePrefix  = "arXiv",
    primaryClass   = "hep-ex",
}

@Article{DAPR-2018-01,
    author         = "{ATLAS Collaboration}",
    title          = "{ATLAS data quality operations and performance for 2015--2018 data-taking}",
    journal        = "JINST",
    volume         = "15",
    year           = "2020",
    pages          = "P04003",
    doi            = "10.1088/1748-0221/15/04/P04003",
    reportNumber   = "CERN-EP-2019-207",
    eprint         = "1911.04632",
    archivePrefix  = "arXiv",
    primaryClass   = "physics.ins-det",
}

@Article{EGAM-2018-01,
    author         = "{ATLAS Collaboration}",
    title          = "{Electron and photon performance measurements with the ATLAS detector using the 2015--2017 LHC proton--proton collision data}",
    journal        = "JINST",
    volume         = "14",
    year           = "2019",
    pages          = "P12006",
    doi            = "10.1088/1748-0221/14/12/P12006",
    reportNumber   = "CERN-EP-2019-145",
    eprint         = "1908.00005",
    archivePrefix  = "arXiv",
    primaryClass   = "hep-ex",
}

@Article{EXOT-2018-08,
    author         = "{ATLAS Collaboration}",
    title          = "{Search for high-mass dilepton resonances using \(139\,\text{fb}^{-1}\) of \(pp\) collision data collected at \(\sqrt{s} = 13\,\text{TeV}\) with the ATLAS detector}",
    journal        = "Phys. Lett. B",
    volume         = "796",
    year           = "2019",
    pages          = "68",
    doi            = "10.1016/j.physletb.2019.07.016",
    reportNumber   = "CERN-EP-2019-030",
    eprint         = "1903.06248",
    archivePrefix  = "arXiv",
    primaryClass   = "hep-ex",
}

@Article{HIGG-2018-27,
    author         = "{ATLAS Collaboration}",
    title          = "{Search for resonances decaying into photon pairs in \(139\,\text{fb}^{-1}\) of \(pp\) collisions at \(\sqrt{s} = 13\,\text{TeV}\) with the ATLAS detector}",
    journal        = "Phys. Lett. B",
    volume         = "822",
    year           = "2021",
    pages          = "136651",
    doi            = "10.1016/j.physletb.2021.136651",
    reportNumber   = "CERN-EP-2020-248",
    eprint         = "2102.13405",
    archivePrefix  = "arXiv",
    primaryClass   = "hep-ex",
}

@Article{PIX-2018-001,
    author         = "Abbott, B. and others",
    title          = "{Production and integration of the ATLAS Insertable B-Layer}",
    journal        = "JINST",
    volume         = "13",
    year           = "2018",
    pages          = "T05008",
    doi            = "10.1088/1748-0221/13/05/T05008",
    eprint         = "1803.00844",
    archivePrefix  = "arXiv",
    primaryClass   = "physics.ins-det",
}

@Article{TRIG-2018-05,
    author         = "{ATLAS Collaboration}",
    title          = "{Performance of electron and photon triggers in ATLAS during LHC Run~2}",
    journal        = "Eur. Phys. J. C",
    volume         = "80",
    year           = "2020",
    pages          = "47",
    doi            = "10.1140/epjc/s10052-019-7500-2",
    reportNumber   = "CERN-EP-2019-169",
    eprint         = "1909.00761",
    archivePrefix  = "arXiv",
    primaryClass   = "hep-ex",
}

@Article{EXOT-2019-16,
    author         = "{ATLAS Collaboration}",
    title          = "{Search for new non-resonant phenomena in high-mass dilepton final states with the ATLAS detector}",
    journal        = "JHEP",
    volume         = "11",
    year           = "2020",
    pages          = "005",
    doi            = "10.1007/JHEP11(2020)005",
    reportNumber   = "CERN-EP-2020-066",
    eprint         = "2006.12946",
    archivePrefix  = "arXiv",
    primaryClass   = "hep-ex",
    related        = "EXOT-2019-16-err",
    relatedstring  = "Erratum:",
}

@Article{EXOT-2019-40,
    author         = "{ATLAS Collaboration}",
    title          = "{Search for periodic signals in the dielectron and diphoton invariant mass spectra using \(139\,\text{fb}^{-1}\) of \(pp\) collisions at \(\sqrt{s} = 13\,\text{TeV}\) with the ATLAS detector}",
    year           = "2023",
    reportNumber   = "CERN-EP-2023-073",
    eprint         = "2305.10894",
    archivePrefix  = "arXiv",
    primaryClass   = "hep-ex",
}

@Booklet{ATL-SOFT-PUB-2021-001,
    author         = "{ATLAS Collaboration}",
    title          = "{The ATLAS Collaboration Software and Firmware}",
    howpublished   = "{ATL-SOFT-PUB-2021-001}",
    url            = "https://cds.cern.ch/record/2767187",
    year           = "2021",
}

@Booklet{ATL-SOFT-PUB-2021-003,
    author         = "{ATLAS Collaboration}",
    title          = "{ATLAS Computing Acknowledgements}",
    howpublished   = "{ATL-SOFT-PUB-2021-003}",
    url            = "https://cds.cern.ch/record/2776662",
    year           = "2021",
}

@Report{ATLAS-TDR-2010-19,
    author         = "{ATLAS Collaboration}",
    title          = "{ATLAS Insertable B-Layer Technical Design Report}",
    type           = "ATLAS-TDR-19; CERN-LHCC-2010-013",
    year           = "2010",
    url            = "https://cds.cern.ch/record/1291633",
    related        = "ATLAS-TDR-2010-19-addm",
    relatedstring  = "Addendum:",
}

@Article{CMS-EXO-17-017,
    author         = "{CMS Collaboration}",
    title          = "{Search for physics beyond the standard model in high-mass diphoton events from proton--proton collisions at \(\sqrt{s} = 13\,\text{TeV}\)}",
    journal        = "Phys. Rev. D",
    volume         = "98",
    year           = "2018",
    pages          = "092001",
    doi            = "10.1103/PhysRevD.98.092001",
    reportNumber   = "CERN-EP-2018-219",
    eprint         = "1809.00327",
    archivePrefix  = "arXiv",
    primaryClass   = "hep-ex",
}

@Article{CMS-EXO-19-019,
    author         = "{CMS Collaboration}",
    title          = "{Search for resonant and nonresonant new phenomena in high-mass dilepton final states at \(\sqrt{s} = 13\,\text{TeV}\)}",
    journal        = "JHEP",
    volume         = "07",
    year           = "2021",
    pages          = "208",
    doi            = "10.1007/JHEP07(2021)208",
    reportNumber   = "CERN-EP-2021-026",
    eprint         = "2103.02708",
    archivePrefix  = "arXiv",
    primaryClass   = "hep-ex",
}

@Booklet{ATLAS-CONF-2019-021,
    author         = "{ATLAS Collaboration}",
    title          = "{Luminosity determination in \(pp\) collisions at \(\sqrt{s} = 13\,\text{TeV}\) using the ATLAS detector at the LHC}",
    howpublished   = "{ATLAS-CONF-2019-021}",
    url            = "https://cds.cern.ch/record/2677054",
    year           = "2019",
}

@Booklet{ATL-PHYS-PUB-2010-013,
    author         = "{ATLAS Collaboration}",
    title          = "{The simulation principle and performance of the ATLAS fast calorimeter simulation FastCaloSim}",
    howpublished   = "{ATL-PHYS-PUB-2010-013}",
    url            = "https://cds.cern.ch/record/1300517",
    year           = "2010",
}

@Booklet{ATL-PHYS-PUB-2014-021,
    author         = "{ATLAS Collaboration}",
    title          = "{ATLAS Pythia~8 tunes to \(7~\text{TeV}\) data}",
    howpublished   = "{ATL-PHYS-PUB-2014-021}",
    url            = "https://cds.cern.ch/record/1966419",
    year           = "2014",
}

@Booklet{ATL-PHYS-PUB-2016-002,
    author         = "{ATLAS Collaboration}",
    title          = "{Multi-boson simulation for \(13~\text{TeV}\) ATLAS analyses}",
    howpublished   = "{ATL-PHYS-PUB-2016-002}",
    url            = "https://cds.cern.ch/record/2119986",
    year           = "2016",
}

@Booklet{ATL-PHYS-PUB-2016-003,
    author         = "{ATLAS Collaboration}",
    title          = "{Monte Carlo Generators for the Production of a \(W\) or \(Z/\gamma^*\) Boson in Association with Jets at ATLAS in Run 2}",
    howpublished   = "{ATL-PHYS-PUB-2016-003}",
    url            = "https://cds.cern.ch/record/2120133",
    year           = "2016",
}

@Booklet{ATL-PHYS-PUB-2016-017,
    author         = "{ATLAS Collaboration}",
    title          = "{The Pythia~8 A3 tune description of ATLAS minimum bias and inelastic measurements incorporating the Donnachie--Landshoff diffractive model}",
    howpublished   = "{ATL-PHYS-PUB-2016-017}",
    url            = "https://cds.cern.ch/record/2206965",
    year           = "2016",
}

@Booklet{ATL-PHYS-PUB-2017-022,
    author         = "{ATLAS Collaboration}",
    title          = "{Electron and photon reconstruction and performance in ATLAS using a dynamical, topological cell clustering-based approach}",
    howpublished   = "{ATL-PHYS-PUB-2017-022}",
    url            = "https://cds.cern.ch/record/2298955",
    year           = "2017",
}
 
\clearpage
 
\begin{flushleft}
\hypersetup{urlcolor=black}
{\Large The ATLAS Collaboration}

\bigskip

\AtlasOrcid[0000-0002-6665-4934]{G.~Aad}$^\textrm{\scriptsize 103}$,
\AtlasOrcid[0000-0002-5888-2734]{B.~Abbott}$^\textrm{\scriptsize 121}$,
\AtlasOrcid[0000-0002-1002-1652]{K.~Abeling}$^\textrm{\scriptsize 55}$,
\AtlasOrcid[0000-0001-5763-2760]{N.J.~Abicht}$^\textrm{\scriptsize 49}$,
\AtlasOrcid[0000-0002-8496-9294]{S.H.~Abidi}$^\textrm{\scriptsize 29}$,
\AtlasOrcid[0000-0002-9987-2292]{A.~Aboulhorma}$^\textrm{\scriptsize 35e}$,
\AtlasOrcid[0000-0001-5329-6640]{H.~Abramowicz}$^\textrm{\scriptsize 152}$,
\AtlasOrcid[0000-0002-1599-2896]{H.~Abreu}$^\textrm{\scriptsize 151}$,
\AtlasOrcid[0000-0003-0403-3697]{Y.~Abulaiti}$^\textrm{\scriptsize 118}$,
\AtlasOrcid[0000-0003-0762-7204]{A.C.~Abusleme~Hoffman}$^\textrm{\scriptsize 138a}$
% ...
\bigskip
\\

% ...
$^{29}$Physics Department, Brookhaven National Laboratory, Upton NY; United States of America.\\
% ...
$^{35}$$^{(a)}$Facult\'e des Sciences Ain Chock, R\'eseau Universitaire de Physique des Hautes Energies - Universit\'e Hassan II, Casablanca;$^{(b)}$Facult\'{e} des Sciences, Universit\'{e} Ibn-Tofail, K\'{e}nitra;$^{(c)}$Facult\'e des Sciences Semlalia, Universit\'e Cadi Ayyad, LPHEA-Marrakech;$^{(d)}$LPMR, Facult\'e des Sciences, Universit\'e Mohamed Premier, Oujda;$^{(e)}$Facult\'e des sciences, Universit\'e Mohammed V, Rabat;$^{(f)}$Institute of Applied Physics, Mohammed VI Polytechnic University, Ben Guerir; Morocco.\\
% ...
$^{49}$Fakult\"{a}t Physik , Technische Universit{\"a}t Dortmund, Dortmund; Germany.\\
% ...
$^{55}$II. Physikalisches Institut, Georg-August-Universit\"{a}t G\"ottingen, G\"ottingen; Germany.\\
% ...
$^{103}$CPPM, Aix-Marseille Universit\'e, CNRS/IN2P3, Marseille; France.\\
% ...
$^{118}$Department of Physics, New York University, New York NY; United States of America.\\
% ...
$^{121}$Homer L. Dodge Department of Physics and Astronomy, University of Oklahoma, Norman OK; United States of America.\\
% ...
$^{138}$$^{(a)}$Departamento de F\'isica, Pontificia Universidad Cat\'olica de Chile, Santiago;$^{(b)}$Millennium Institute for Subatomic physics at high energy frontier (SAPHIR), Santiago;$^{(c)}$Instituto de Investigaci\'on Multidisciplinario en Ciencia y Tecnolog\'ia, y Departamento de F\'isica, Universidad de La Serena;$^{(d)}$Universidad Andres Bello, Department of Physics, Santiago;$^{(e)}$Instituto de Alta Investigaci\'on, Universidad de Tarapac\'a, Arica;$^{(f)}$Departamento de F\'isica, Universidad T\'ecnica Federico Santa Mar\'ia, Valpara\'iso; Chile.\\
% ...
$^{151}$Department of Physics, Technion, Israel Institute of Technology, Haifa; Israel.\\
$^{152}$Raymond and Beverly Sackler School of Physics and Astronomy, Tel Aviv University, Tel Aviv; Israel.\\
% ...
$^{a}$ Also Affiliated with an institute covered by a cooperation agreement with CERN.\\
$^{b}$ Also at An-Najah National University, Nablus; Palestine.\\
$^{c}$ Also at Borough of Manhattan Community College, City University of New York, New York NY; United States of America.\\
$^{d}$ Also at Center for High Energy Physics, Peking University; China.\\
$^{e}$ Also at Center for Interdisciplinary Research and Innovation (CIRI-AUTH), Thessaloniki; Greece.\\
$^{f}$ Also at Centro Studi e Ricerche Enrico Fermi; Italy.\\
% ...

\end{flushleft}


\clearpage

\end{document}